\documentclass[prc,twocolumn,aps,reprint,nofootinbib,showpacs,superscriptaddress]{revtex4-1}

\usepackage{graphicx}
\usepackage{natbib}
\usepackage{amssymb,amsmath,amstext,amsthm,amsfonts}

\usepackage[usenames]{color}

\newcommand{\GeV}{\; \mathrm{GeV}}
\newcommand{\MeV}{\; \mathrm{MeV}}

\newcommand{\dd}{\mathrm{d}}

\begin{document}

\title{Many-Body Interactions of Neutrinos with Nuclei - Observables}
\author{O. Lalakulich}
\affiliation{Institut f\"ur Theoretische Physik, Universit\"at Giessen, Germany}
\author{K. Gallmeister}
\affiliation{Institut f\"ur Theoretische Physik, Johann Wolfgang Goethe-Universit\"at Frankfurt, Germany}
\author{U. Mosel}
\email{mosel@physik.uni-giessen.de}
\affiliation{Institut f\"ur Theoretische Physik, Universit\"at Giessen, Germany}

\begin{abstract}
\begin{description} \item[Background] The total inclusive cross sections obtained for quasielastic (QE) scattering in the Mini Booster Neutrino Experiment (MiniBooNE) are significantly larger than those calculated by all models based on the impulse approximation and using the world average value for the axial mass of $M_A \approx 1 \GeV$. This discrepancy has led to various, quite different explanations in terms of increased axial masses, changes in the functional form of the axial form factor, increased vector strength in nuclei, and initial two-particle interactions. This is disconcerting since the neutrino energy reconstruction depends on the reaction mechanism. \item[Purpose] We investigate whether exclusive observables, such as nucleon knock-out, can be used to distinguish between the various proposed reaction mechanisms. We determine the influence of 2p-2h excitations on the neutrino energy reconstruction. \item[Method] We extend the Giessen Boltzmann--Uehling--Uhlenbeck (GiBUU) model by explicitly incorporating initial 2p-2h excitations.  \item[Results]We calculate inclusive cross sections and numbers and spectra of knock-out nucleons and show their sensitivity to the presence of 2p-2h initial excitations. We also discuss the influence of 2p-2h excitations on the neutrino energy reconstruction. \item[Conclusions] Inclusive double-differential cross sections, depending only on muon variables, are fairly insensitive to the reaction mechanism. 2p-2h excitations lead to an increase in the number $n$ of knock-out nucleons for $n \ge 2$, while only $n=1$ knock-out remains a clean signal of true QE scattering. The spectra of knock-out nucleons are also changed, but their shape is hardly affected. In the energy reconstruction, 2p-2h interactions as well as $\Delta$ excitations lead to a downward shift of the reconstructed energy; this effect 2p-2h excitations disappears at higher energies because the 2p-2h influence is spread out over a wider energy range.
\end{description}
\end{abstract}

\maketitle
\section{Introduction}

One of the essential ingredients for any extraction of the neutrino masses from oscillation experiments is the neutrino energy. This energy is not known a priori in present-day experiments, because neutrino beams are quite broad in energy due to their production mechanisms. While at higher energies calorimetric methods may play a role, at lower energies (a few hundred MeV to a few GeV) quasielastic (QE) scattering has been used to determine the incoming neutrino energy on an event-by-event basis. This method relies on an identification of the reaction mechanism (interaction of the neutrino with a single nucleon). It also relies on the use of quasifree kinematics that describes neutrino scattering on a single, free nucleon at rest, thus neglecting  any Fermi-motion effects; binding is taken
into account only by a constant removal energy.

In theoretical calculations the QE cross section is determined by an interplay of vector and axial couplings with their corresponding form factors. The vector couplings can be rather well determined from electron scattering experiments on the nucleon that work at a fixed energy and permit one to determine the relevant kinematic parameters, i.e.\ energy and momentum transfer, in each event. The corresponding form factors have been shown to have a complicated, non-dipole form \cite{Arrington:2006zm}. For the axial couplings the situation is less well determined. Here the data come from electro-pion production and older neutrino data on the nucleon or deuterium with large uncertainties. They have been analyzed by making a dipole ansatz and then extracting the axial mass from a fit to data. The world average for the axial mass parameter is found to be $M_A = 1.026 \GeV$ \cite{Bernard:2001rs}.

It came, therefore, as a surprise when the Mini Booster Neutrino Experiment (MiniBooNE) at Fermilab published its results on QE scattering. The analyses of both charged current (CC) and neutral current (NC) high-statistics QE events showed a clear excess of cross section over that expected for QE scattering from a Fermi-gas model \cite{:2007ru,AguilarArevalo:2010zc,AguilarArevalo:2010cx}. A similar result had been obtained by the K2K experiment \cite{Gran:2006jn} that worked with a neutrino flux peaked at a slightly higher neutrino energy ($\approx 1 \GeV$) than the MiniBooNE experiment ($\approx 0.7 \GeV$). In both cases the flux distributions are rather broad and have a considerable overlap. In contrast, the NOMAD experiment working at significantly higher energies (between about 5 and $100\GeV$) observed no such excess of measured over expected quasielastic cross section \cite{Lyubushkin:2008pe}.
Both the MiniBooNE and the K2K experiments could obtain good fits to their data in the Fermi-gas model only when the axial mass was considerably increased to $M_A = 1.23 \GeV$ \cite{:2007ru} or even $M_A = 1.35 \GeV$; the latter was obtained from a shape-only fit to $\dd\sigma/\dd Q^2$ \cite{AguilarArevalo:2010zc}. The sizable increase in the axial mass needed to describe the data cannot be ascribed to deficiencies in the Fermi-gas model alone. Indeed, Benhar et al.\ showed that a model based on state-of-the-art nucleon spectral functions required an even larger axial mass ($M_A = 1.6 \GeV$) for a fit of the differential data \cite{Benhar:2010nx}.

In contrast to electron scattering, the energy of the incoming neutrino is not fixed due to the broad energy profile of the neutrino beam; only energy and angle of the outgoing lepton can be determined. The cross section at fixed energy and scattering angle of the outgoing lepton then picks up
contributions from different kinematical regions and thus quite different processes \cite{Benhar:2011ef}.
For example, in the MiniBooNE energy regime there is a strong entanglement of true QE scattering and pion production \cite{Leitner:2010kp}. Events in which a pion produced in the initial interaction is absorbed while traveling through the nucleus can also lead to knock-out nucleons without any pions and thus look very similar to genuine QE events. In Refs.\ \cite{Leitner:2010kp,Leitner:2010jv} we have shown that these events can affect the energy reconstruction and the extraction of oscillation parameters from neutrino flux comparisons. However, in the data mentioned these types of events had already been removed with the help of event generators. While this removal may not be perfect, the inaccuracies connected with it are probably not large enough to explain the observed excess.

Here it is interesting to note that also the explicit pion cross sections obtained by MiniBooNE are by about a factor 1.5 - 2 larger than those obtained with the very same event generator (NUANCE) used to analyze the MiniBooNE data \cite{AguilarArevalo:2010zc}. A similar result ($>50\%$ excess) holds for simulations with the theory-based generator GiBUU. In Ref.~\cite{Leitner:2009de,Lalakulich:2011ne} we have discussed that the pion yields measured by the MiniBooNE experiment \cite{AguilarArevalo:2010xt,AguilarArevalo:2010bm} considerably exceed the pion yields calculated in the impulse approximation; the data agree essentially with the calculations before any of the strong pion-nucleus final-state interactions (FSI) are taken into account.\footnote{Qualitatively the same result is found in calculations with the neutrino event generator NUANCE \cite{AguilarArevalo:2010xt,AguilarArevalo:2010bm}.} The most obvious explanation for the common excess in QE scattering \emph{and} pion production could be an inaccuracy in the flux determination: a 30\% higher flux would bring both data sets in much better agreement with theory.\footnote{The shape of $\dd\sigma/\dd Q^2$ could be explained by RPA correlations \cite{AlvarezRuso:2009ad}.} The MiniBooNE experiment, however, gives a possible inaccuracy of only about 10\% for the flux \cite{AguilarArevalo:2008yp}, so that this explanation seems to be -- at least partly -- ruled out.

One major difference of the older data on elementary targets is that the mentioned experiments use nuclei as target material; in the case of the MiniBooNE experiment this is oil (CH$_2$), while for K2K it was water (H$_2$O). It is, therefore, tempting to assume that the observed effect is due to some nuclear in-medium effect on the axial current. Both the analyses of MiniBooNE and K2K changed only the axial coupling, leaving the vector coupling unchanged. In a similar class of models belongs an analysis in which the functional form of the axial form factor was fitted to the MiniBooNE data, i.e.,\ the dipole ansatz for the axial form factor was no longer made. This leads to an axial mass $M_A = 0.85 $-$ 1 \GeV$ (defined via the derivative of the formfactor at $Q^2 = 0 \GeV^2$) \cite{Bhattacharya:2011ah}. In contrast, a recent explanation of the observed excess by Bodek et al.\ \cite{Bodek:2011ps} is based on the observation that the transverse response of nuclei in inclusive electron scattering is underestimated by the Fermi-gas model. By fitting a nuclear transverse-enhancement factor for the electron-data, i.e.,\ by changing only the vector current, the authors of Ref.\ \cite{Bodek:2011ps} then are able to describe the MiniBooNE data vs.\ reconstructed energy and even - approximately - the disappearance of the excess in the NOMAD experiment. It has recently been shown \cite{Sobczyk:2012ah} that this ad-hoc ansatz of \cite{Bodek:2011ps} also describes the double-differential cross sections.
Finally, Meucci et al.\ \cite{Meucci:2011vd} find  within the relativistic Green's function model (RGF), which is basically related to the (one-particle) impulse approximation, a very strong effect of FSI. These authors are able to describe the MiniBooNE data perfectly well when including the FSI between the ejected nucleon and the residual nucleus. In this case the FSI contain implicitly effects of other reaction mechanisms. This is similar in spirit to the work of Ref.\ \cite{Bleve:2000hc} where the authors absorbed higher excitations beyond 1p-1h into the FSI by using Feshbach's projection formalism.

All the explanations described in the preceding paragraph rely on the so-called impulse approximation in which the incoming neutrino interacts with one nucleon at a time  only. Early work on electron scattering had also shown the importance of random-phase approximation (RPA) correlations in the QE-peak region, where these correlations tend to lower the cross section \cite{Alberico:1983zg}; Kim et al.\ showed that similar effects also appear in neutrino-induced reactions \cite{Kim:1994zea}. Furthermore, from electron scattering we know that also more complicated processes take place, in which the incoming electron interacts with two nucleons at the same time. Indeed, in inclusive inelastic electron scattering on nuclei these so-called two-particle--two-hole (2p-2h) processes become significant at larger energy transfers beyond the QE peak, in the so-called dip region between the QE peak and the $\Delta$ peak and under the $\Delta$ resonance where the strength could not be explained in a one-particle picture alone. A rather complete summary of early attempts for electron scattering along these lines can be found in Chapter 5 of Ref.\cite{Boffi:1994}. Later work concentrated on a correct relativistic treatment of meson exchange currents (MECs) the dip region, where these indeed lead to a marked contribution to the transverse response \cite{Dekker:1994yc}.

As a consequence of the experimental method to identify QE events in the MiniBooNE experiment (knock-out nucleons are not observed) the measured cross section for neutrinos can indeed also contain contributions from 2p-2h excitations. Delorme and Ericson noticed already in the context of old bubble chamber experiments that two-particle--two-hole excitations could contribute to the total nuclear response \cite{Delorme:1985ps}. Following this suggestion Marteau \cite{Marteau:1999kt}, in analyzing earlier neutrino experiments, included 2p-2h excitations and the RPA in his analysis. At about this time also Bleve et al.\ used a similar ansatz in their study of the nuclear response to neutrinos in the QE region \cite{Bleve:2000hc}. Martini et al.~\cite{Martini:2009uj} were the first to realize that also the MiniBooNE experiment could not separate out the QE process, because the experiment is insensitive to any outgoing nucleons and that 2p-2h processes could thus contribute to the measured quasielastic-like cross section. By combining the RPA  with a calculation of 2p-2h contributions these authors obtain a good description of the MiniBooNE data \cite{Martini:2009uj,Martini:2010ex,Martini:2011wp}. As expected, the RPA correlations have most effect at forward angles where the squared four-momentum transfer $Q^2$ to the nucleus is small. They die out with increasing angle and with decreasing muon energy, i.e.,\ increasing energy transfer. The calculations do not describe the falloff of the QE excess at the higher NOMAD energies, but they make predictions for the antineutrino cross sections; for these the 2p-2h contributions are significantly smaller than for the neutrino ones.

The idea of Martini et al.\ has been taken up by various authors who try to improve on the detailed theoretical ingredients. Nieves et al.\  have obtained in Ref.\ \cite{Nieves:2011pp} a very good agreement with the energy-separated data, but the QE contribution shown there relies on non-relativistic approximations which are not reliable at the higher neutrino energies. This can also be seen by comparing their QE cross section $\sigma(E_\nu)$ with the results of other models in Ref.~\cite{Boyd:2009zz}; a corrected result gives still a good, but somewhat less perfect agreeement for these data with a clear discrepancy towards the lower neutrino energies \cite{Nieves-Vicente:2011}. In another calculation by the same authors that employs the correct relativistic corrections, but leaves out any final state interactions (FSI) \cite{Nieves:2011yp}, these authors have obtained a very good description of the measured inclusive double-differential cross sections, when simultaneously readjusting the experimental flux within its inaccuracies. Their calculations build on earlier work on electron-induced reactions \cite{Gil:1997bm,Gil:1997jg} and provide a consistent theoretical framework; the corresponding electron data, though mostly at somewhat lower energies than relevant for the neutrino experiments, are described quite well. As in the work of Martini et al.\ the disappearance of the excess when going up to the NOMAD energy regime cannot be studied. For antineutrinos these authors predict, contrary to the results of Martini et al., a sizable effect of 2p-2h excitations so that the antineutrino cross section becomes nearly as large as that for neutrinos. The FSI effects, which are neglected in Ref.~\cite{Nieves:2011yp}, are estimated to be small ($< 7 \%$) for the total cross section, but could be larger for the lower neutrino energies. The fact that Meucci et al.\ \cite{Meucci:2011vd} obtain a good description of the energy-separated MiniBooNE data due to strong FSI within an RGF calculation just underlines the fact that different theoretical ingredients may lead to the same final agreement with the flux-averaged neutrino data.

Finally, an interesting approach is that by Amaro et al.\ \cite{Amaro:2010sd,Amaro:2011qb} which starts with a phenomenological model for the neutrino interactions with nuclei that is based on the superscaling (SUSA) behavior of electron scattering data. The scaling function thus determined is then directly taken over to neutrino interactions with some 2p-2h contributions due to meson exchange currents added; those are so far neither gauge invariant nor do they contain the axial contributions. There are no RPA correlations explicitly taken into account, but they may be contained in the scaling function. The overall effect of using this model is an increase of the inclusive cross section for neutrinos; at forward muon angles the calculations come close to the data, but the MEC contributions die out fast with increasing angle so that the cross section is significantly underestimated at backward angles. As a consequence the energy-separated cross section obtained for the MiniBooNE experiment -- while being higher than that obtained from SUSA alone -- still underestimates the experimental result even when 2p-2h contributions are added. Recently, a strong difference between neutrino and antineutrino cross sections has been obtained within this model, with the 2p-2h effects being significantly larger for antineutrinos than for neutrinos \cite{Amaro:2011aa}. The latter result is in contradiction to that obtained by Martini et al.\ and by Nieves et al.  For short, but rather comprehensive discussions of 2p-2h effects see also Refs.~\cite{Martini:2011ui,Alvarez-Ruso:2011}.

In summary, there exist various, mutually exclusive explanations of the observed QE excess. The 1p-1h models disagree in attributing the observed effects either to the axial coupling \cite{:2007ru,Bhattacharya:2011ah}, the vector coupling \cite{Bodek:2011ps}, or to final-state interactions \cite{Meucci:2011vd}. Alternatively -- or in addition -- two-body interaction effects, that are present in electron scattering most likely also play a role in neutrino experiments. Various models agree in their importance, but still suffer from problems connected with their treatment of gauge invariance, final-state interactions, uncertainties in relativistic corrections \cite{Amaro:2011qb,Martini:2011wp,Nieves:2011yp}, and upper energy limits for their applicability \cite{Nieves:2011yp}. As a consequence, the precise strength of 2p-2h processes and their energy dependence in neutrino experiments is still uncertain.

We think that progress can be made by following either one of two different paths:
\begin{itemize}
 \item On the theoretical side, the calculations of 2p-2h contributions can be improved so that gauge-invariant calculations of nuclear correlations and meson exchange currents become available and RPA effects and final state interactions are consistently calculated with the same forces. At the end, for any comparison with experiment, the calculated cross sections have to be averaged over quite broad energy distributions thus possibly wiping out many details of the interaction matrix elements.

 \item A more phenomenological approach suggests itself in which detailed calculations of 2p-2h contributions are avoided by starting with a flux-averaged matrix element. Such an approach, if incorporated into an event generator, can lead  faster to nevertheless reliable results on experimental observables beyond the inclusive cross sections.
\end{itemize}

While theoretical consistency arguments may favor one model over others the ultimate answer can come only from experiments. Inclusive measurements are obviously not very sensitive to details of the reaction mechanism because they always represent sums over various reaction mechanisms. For neutrinos this is even more so because experiments always involve an average over incoming energies. A clarification of the reaction mechanism must, therefore, come from experiments other than just inclusive reactions; exclusive or semi-inclusive reactions can give more specific information.
If one wants to go beyond inclusive cross sections both of the approaches named above must be combined with an event generator that contains a very good, well tested treatment of FSI. In addition, the data necessarily always involve also some pion degrees of freedom, even if no final-state pions are observed \cite{Leitner:2010kp}. Because of the broad energy distributions in the neutrino beams one thus has to develop and use methods to describe consistently scattering processes at low \emph{and} high momentum transfers with different reaction mechanisms (1p-1h, 2p-2h, QE, resonance excitation, and deep inelastic scattering (DIS)) because all of these can contribute to the measured cross sections \cite{Ankowski:2010yh}.

In this paper we follow the second path from the two named above. For the main question raised here (1p-1h vs\ 2p-2h) it is obvious to try to explore possible observable consequences of the presence of two-body interactions in the numbers and spectra of knock-out nucleons. We will, therefore, investigate the impact of 2p-2h processes on knock-out nucleons in a simple model that avoids the quite significant difficulties of calculating the MEC and correlation effects by making a physically motivated ansatz for the relevant matrix element. The model aims for a first exploration of what will happen if two-body interactions play a noticeable role; in a later stage we could then implement the results of sophisticated calculations. We also analyze the impact of 2p-2h processes on the reconstruction of the neutrino energy.

While in this paper we deal only with charged current (CC) events it is obvious that such studies are most relevant for neutral current (NC) interactions  where only outgoing hadrons are available for the energy and $Q^2$ reconstruction.

\section{Model for 2p-2h contributions}
\subsection{Basic properties of GiBUU}
The starting point for our studies is the GiBUU (Giessen Boltzmann-Uehling-Uhlenbeck) model that has been developed originally as an event generator for heavy-ion reactions and then extended to descriptions of reactions involving elementary projectiles impinging on nuclei. For all the various applications, the very same physics input and code are used; this sets the GiBUU model apart from many other event generators. A complete description of the GiBUU model can be found in a recent review \cite{Buss:2011mx} which also contains a complete list of references to applications of the GiBUU model and a discussion of results obtained with it.

In the GiBUU model the spectral one-particle phase-space distributions, $F(x,p)$, of all particles are obtained by solving the coupled Kadanoff-Baym equations
\cite{Kad-Baym:1962} for each particle species in their gradient-expanded form \cite{Botermans:1990qi}
\begin{equation} \label{eq:os-transp.9}
\mathcal{D} F(x,p) - \text{tr} \left\{ {\Gamma f},{\Re S^{\text{ret}}(x,p)}\right\}_{\rm pb} = C(x,p)~,
\end{equation}
with
\begin{equation}
\mathcal{D} F = \left\{p_0 - H,F\right\}_{\rm pb}~.
\end{equation}
Here $\{\ldots\}_{\rm pb}$ denotes a Poisson bracket.
In the so-called backflow term [second term on the left-hand side in Eq.~(\ref{eq:os-transp.9})], which is essential for off-shell transport,  $f(x,p)$ is the phase-space density related to $F$ by
\begin{equation}
F(x,p) = 2 \pi g f(x,p) A(x,p) ~,
\end{equation}
where $A(x,p)$ is the spectral function of the particle\footnote{$A$ is normalized as $\int_0^\infty A(x,p) \dd p_0 = 1$.} and $g$ is the spin-degeneracy factor. For on-shell particles [$A = \delta(p^0 - E)$] the phase-space density $f$ is connected to the nuclear density by
\begin{equation}    \label{rho}
\rho(x) = g \int \frac{\dd^3p}{(2\pi)^3} f(x,p) ~.
\end{equation}
The quantity $\Gamma$ in the backflow term is the width of the spectral function,
and $S^{\text{ret}}(x,p)$ denotes the retarded Green's function.  Off-shell transport is thus included and leads to the correct asymptotic spectral functions of particles when they leave the nucleus. The expression $C(x,p)$ on the right-hand side of Eq.~(\ref{eq:os-transp.9}) denotes the collision term that couples all particle species; it contains both a gain and a loss term. For a short derivation of this transport equation and further details we refer the reader to Ref.~\cite{Buss:2011mx}.

Because the GiBUU model was developed to be used for many different reaction mechanisms, it provides a consistent framework for the description of broad-energy beam neutrino experiments. The GiBUU model has undergone extensive testing with various reaction types (see, e.g.,\ Refs.~\cite{Buss:2011mx,Leitner:2009ke}); for the purpose here its validation with photo-produced mesons on nuclei is most relevant. In Refs.~\cite{Leitner:2006ww,Leitner:2006sp} we have presented the first applications of the model to neutrino reactions with nuclei and have analyzed both QE scattering and pion production. The calculations include the FSI of all particles; furthermore we use relativistic kinematics throughout.
The procedure for QE and pion-production events is as described in Ref.~\cite{Leitner:2006ww}; further details on the GiBUU model can be found in a recent review \cite{Buss:2011mx}. Relevant for the following discussions is that the QE cross section is calculated with an axial mass of $M_A = 1 \GeV$. At the energies relevant for MiniBooNE, pion production proceeds overwhelmingly through the $\Delta$ resonance; its coupling strength is given by partial conservation of the axial current (PCAC) and its form factor is described by a modified dipole form factor, as explained in Refs.~\cite{Leitner:2006ww,Leitner:2006sp}.

For the MiniBooNE data discussed here, the model, which was so far based on the impulse approximation and did not contain any two-body interactions for the incoming neutrino, also -- as all the other models \cite{Boyd:2009zz} -- underestimates the measured QE cross sections.
We have now extended this model by including such processes so that now a generator is available that contains the 2p-2h effects. The extension will be described in the next subsection.

\subsection{Inclusion of two-nucleon interactions}
In Ref.~\cite{Leitner:2006ww} we have described how we treat the CC neutrino interactions with nuclei within the impulse approximation. Here we now develop the relevant expressions for an extension to 2p-2h processes. The starting point is the triple-differential cross section for the reaction $\nu(k) + A \to l^-(k') + X$
\begin{equation}    \label{sigma}
\frac{\dd^3\sigma}{\dd\Omega'\, \dd E'} = \frac{|\mathbf{k'}|}{E_\nu} \frac{G^2}{4\pi^2} \left|\mathcal{T} \right|^2 ~,
\end{equation}
where $k=(E_\nu,\mathbf{k})$ and $k'=(E',\mathbf{k}')$ are the incoming neutrino and outgoing lepton momenta, respectively, $\Omega'$ is the scattering angle of the outgoing lepton and $G$ is the Fermi constant.
The squared invariant amplitude $|\mathcal{T}|^2 = L_{\mu \lambda} W^{\mu \lambda}$ is given by the contraction of the leptonic tensor $L$ and the nuclear hadronic tensor $W$.

The total cross section for the interaction of a neutrino with a nucleus can be related to the collision rate $\Gamma$ of the neutrino and is given by \cite{Nieves:2004wx}
\begin{equation}    \label{stot}
\sigma_{\rm tot} = \int \frac{\dd^3\sigma}{\dd^3k'}\, \dd^3k' = \frac{E_\nu}{|\mathbf{k}|}\int \Gamma(x,k)  \, \dd^3x
\end{equation}
Here the local density approximation has been used. The collision rate (\ref{stot}) is directly related to the imaginary part of the neutrino self/energy.

In transport theory the collision rate can be obtained from the loss term in the collision term in Eq.~(\ref{eq:os-transp.9}). This term contains one-body (resonance decay, for example), two-body (QE scattering, for example), and three-body (2p-2h) terms. For our purpose here only the two-body and the three-body terms are relevant so that $\Gamma$ is the collision rate for two-body and three-body processes $\Gamma = \Gamma^{(2)} + \Gamma^{(3)}$. The spatial integration in Eq.~(\ref{stot}) extends over the nuclear volume. The two-body interactions contained in $\Gamma^{(2)}$ have been discussed in detail in Refs.~\cite{Leitner:2006ww,Leitner:2006sp} and are already implemented in the GiBUU model. They contain QE scattering and pion production through resonance excitation.

For the three-body collision processes ($\nu + N_1N_2 \to l' + N_3 N_4$) of interest here, the collision rate, which represents the imaginary part of the self energy in local density approximation, is given by (for details and notation see Sec.\ 3.3 in Ref.~\cite{Buss:2011mx})
\begin{widetext}
\begin{eqnarray} \label{C^3}
\Gamma^{(3)}_A(x,k) &=& \frac{C_{\rm loss}^3(x,k)}{F_\nu(x,k)} \\
    &=& \mbox{ }  \frac{1}{2E_\nu} \frac{\mathcal{S}_{12}\mathcal{S}_{34}}{ g_{3}g_{4}g_{l'}} \int\,
    \frac{\dd^4p_{1}}{(2\pi)^4 2p_{1}^0} \int\, \frac{
      \dd^4p_{2}}{(2\pi)^4 2p_{2}^0} \int\, \frac{\dd^4k'}{(2\pi)^4
      2{k'}^0} \int\, \frac{\dd^4p_{3}}{(2\pi)^4 2p_{3}^0}
    \int\, \frac{\dd^4p_{4}}{(2\pi)^4 2p_{4}^0} \nonumber \\
    & & \mbox{ } \times F_{1}(x,p_{1}) F_{2}(x,p_{2}) (2\pi)^4
    \delta^{(4)}\left(k+p_1+p_2- k' - p_3-p_4\right)\,
    \overline{|\mathcal{M}_{\nu 12 \to l'34}|^2} \,\overline{F}_{l'}(x,k') \overline{F}_{3}(x,p_{3}) \overline{F}_{4}(x,p_{4}) ~. \nonumber
\end{eqnarray}
\end{widetext}
Here
$\overline{|\mathcal{M}_{\nu 12 \to l'34}|^2}$ is the in-medium matrix element, squared and averaged over the spin states of the initial particles and summed over those of the final particles.
\footnote{The spinors in $\mathcal{M}$ are normalized according to $u^\dagger u = 2E$.}
The $F_i = F(x_i,p_i)$ are the spectral phase-space densities of the two nucleons with which the interaction of the $W$ boson takes place (for $i =1,2$) or of the outgoing nucleons ($i=3,4$), and $\overline{F}(x,p) = 2 \pi g A(x,p)[1 - f(x,p)]$ contains the Pauli blocking. Since we consider a stationary nuclear target all these distributions do not depend on time. The four-momenta of the nucleons are given by $p_i = (p^0_i,\mathbf{p}_i)$ and their energies by $E_i(\mathbf{p}_i) = \sqrt{\mathbf{p}_i^2 + M^2} + U(x,\mathbf{p}_i)$ ($i=1,2,3,4$), where $M$ is the nucleon mass and $U$ is a space- and momentum-dependent nuclear mean-field potential. The $g_i$ are spin-degeneracy factors and the $\mathcal{S}_{ij}$ are symmetry factors ($\mathcal{S}_{ij}=1/2$ for $pp$ or $nn$ pairs, 1 for $pn$ pairs). There is no Pauli blocking for the final lepton so that $f_{l'}$ = 0. Here we neglect all energies and momenta in the recoil nucleus.

The $\delta^{(4)}$ function in Eq.~(\ref{C^3}) limits the degrees of freedom. Since we have three outgoing on-shell particles with in total nine vector components and four energy-momentum conserving constraints, only five degrees of freedom are left. One could thus evaluate cross sections such as
\begin{equation}   \label{d5sdk}
\frac{\dd^5\sigma}{\dd^3k'\,\dd\Omega_3}
\end{equation}
while all the other kinematical quantities, in particular the energies of the two knock-out nucleons, are restricted by energy-momentum conservation. In theoretical studies of photon- and electron-induced two-nucleon knock-out reactions from nuclei, even higher-differential cross sections were evaluated  \cite{Gottfried:1975,Giusti:1991bs,Giusti:1993,VanderSluys:1995rp,Ryckebusch:1996wc}.
For the special case of zero energy and momentum of the residual nucleus they reduce to those calculated here.

In order to separate the initial states from the final states we now define the collision rate, $\Gamma_{NN}$, of a neutrino with two nucleons with momenta $p_1$ and $p_2$ at point $x$ leading to an outgoing lepton with momentum $k'$,
\begin{widetext}
\begin{eqnarray}   \label{Gdef}
\Gamma_{NN}(k,p_1,p_2,k') &=&  \frac{1}{2E_\nu}\, \frac{\mathcal{S}_{34}}{ g_{3}g_{4}} \,\int\, \frac{\dd^4p_{3}}{(2\pi)^4 2p_{3}^0}
    \int\, \frac{\dd^4p_{4}}{(2\pi)^4 2p_{4}^0} \nonumber \\
    & & \mbox{ } \times (2\pi)^4
    \delta^{(4)}\left(k+p_1+p_2- k' - p_3-p_4\right)\,
    \overline{|\mathcal{M}_{\nu 12 \to l'34}|^2} \, \overline{F}_{3}(x,p_{3}) \overline{F}_{4}(x,p_{4}) ~,
\end{eqnarray}
so that we can rewrite Eq.~(\ref{C^3}) into
\begin{eqnarray}  \label{G^3}
\Gamma^{(3)}_A(x,k) &=&    \mathcal{S}_{12} \int\,
    \frac{\dd^4p_{1}}{(2\pi)^4 2p_{1}^0} \int\, \frac{
    \dd^4p_{2}}{(2\pi)^4 2p_{2}^0}  F_{1}(x,p_{1}) F_{2}(x,p_{2})\,\int\, \frac{\dd^4k'}{(2\pi)^4
      2{k'}^0}  \, \Gamma_{NN}(k,p_1,p_2,k') \frac{1}{g_{l'}} \overline{F}_{l'}(x,k') \nonumber \\
    &=&     \mathcal{S}_{12} \int\,
    \frac{\dd^4p_{1}}{(2\pi)^4 2p_{1}^0} \int\, \frac{
    \dd^4p_{2}}{(2\pi)^4 2p_{2}^0}  F_{1}(x,p_{1}) F_{2}(x,p_{2})\,\int\, \frac{\dd^3k'}{(2\pi)^3
      2E'}  \, \Gamma_{NN}(k,p_1,p_2,k') ~.
\end{eqnarray}
In going from the first to the the second line we have used for the outgoing lepton $\overline{F}_{l'} = 2 \pi g_{l'} \delta(k_0' - E')$ so that in Eq.\ (\ref{G^3}) the integration over $k_0'$ can be carried out, reducing the integral over $k'$ to a three-dimensional one. The differential cross section can now be obtained from Eqs.~(\ref{stot}) and (\ref{G^3}),
\begin{equation}      \label{d3sdk}
\frac{\dd^3\sigma^{(3)}}{\dd^3k'} =  \mbox{ }  \frac{E_\nu}{|\mathbf{k}| E'} \frac{1}{2(2\pi)^3}\mathcal{S}_{12} \,
    \int\dd^3x   \int\, \frac{\dd^4p_{1}}{(2\pi)^4 2p_{1}^0} \int\, \frac{\dd^4p_{2}}{(2\pi)^4 2p_{2}^0}\,  F_{1}(x,p_{1}) F_2(x,p_2)  \Gamma_{NN}(k,p_1,p_2,k') ~.
\end{equation}
In this form the actual two-body interaction contained in $\Gamma_{NN}$ is separated from the phase-space distributions of the initial particles. This corresponds to the so-called factorization already used by Gottfried \cite{Gottfried:1975}.

Up to this point the formalism is quite general as far as the outgoing particles are concerned. These could be either $NN$, $N\Delta$, or $\Delta\Delta$ pairs.  We now simplify the two-body collision rate $\Gamma_{NN}$. Since we are primarily interested in nucleons being ejected into the continuum the Pauli-blocking factors $1 - f_{3,4}\ $ can be neglected\footnote{In the actual calculations we keep the Pauli-blocking factors, thus allowing also for a possible recapturing of the knock-out nucleons.} so that $\overline{F} = 2 \pi g A$.

We now eliminate the four-dimensional momentum-conserving $\delta$ function in $\Gamma_{NN}$ by integrating over $\dd^4p_4$. With fixed $P=k-k'+p_1+p_2 = p_3 + p_4$ the function $\Gamma_{NN}$ (\ref{Gdef}) is then given by
\begin{equation}    \label{Gtilde1}
    \Gamma_{NN}(k,p_1,p_2,k') =  \frac{(2\pi)^2}{2 E_\nu} \mathcal{S}_{34} \int\, \frac{\dd^4p_{3}}{(2\pi)^4 2p_{3}^0}
    \frac{1}{ 2 (P^0 - p_3^0)}
    \overline{|\mathcal{M}_{\nu 12 \to l'34}|^2} A_3(x,p_{3}) A_{4}(x,P - p_{3}) ~.
\end{equation}
The remaining integral over $\dd^4p_3$ can be further simplified by using the so-called quasiparticle approximation, i.e.,\ neglecting the width in the spectral functions of the outgoing particles, i.e.,\ restricting ourselves to outgoing nucleons by setting
\begin{equation}
A_i(x,p) = \delta(p_i^0 - \tilde{E}_i)
\end{equation}
with $\tilde{E}_i = \sqrt{\mathbf{p}_i^2 + M^2}$. After integrating Eq.~(\ref{Gtilde1}) over $\dd p_3^0$ one obtains
\begin{equation}    \label{Gtilde2}
    \Gamma_{NN}(k,p_1,p_2,k') =  \frac{\pi}{E_\nu} \mathcal{S}_{34} \int\, \frac{\dd^3p_{3}}{(2\pi)^3 2\tilde{E}_3 2 \tilde{E}_4} \delta(P_0 - \tilde{E}_3 - \tilde{E}_4) \overline{|\mathcal{M}_{\nu 12 \to l'34}|^2}  ~,
\end{equation}
with $\tilde{E}_3 = \sqrt{\mathbf{p}_3^2 + M^2}$ and $\tilde{E}_4 = \sqrt{(\mathbf{P} - \mathbf{p}_3)^2 + M^2} = P_0 - \tilde{E}_3$.
Integrating out now $|\mathbf{p}_3|$, exploiting the $\delta$ function, gives\footnote{The result is identical to that obtained by using the Jacobian given by Ruiz Simo et al \cite{Simo:2014wka}.}
\begin{eqnarray}    \label{Gtilde3}
\Gamma_{NN}(k,p_1,p_2,k') &=&  \frac{1}{E_\nu} \mathcal{S}_{34} \frac{\pi}{4} \int\, \frac{\dd\Omega_3}{(2\pi)^3} \frac{\tilde{\mathbf{p}}_3^2}{\tilde{E}_3 \tilde{E}_4 } \frac{1}{\left|\frac{\dd(\tilde{E}_3 +\tilde{E}_4)}{\dd \left|\mathbf{p}_3\right|}\right|_{\tilde{p}_3} } \,   \overline{|\mathcal{M}_{\nu 12 \to l'34}|^2} \nonumber \\
&=&  \frac{1}{E_\nu} \mathcal{S}_{34} \frac{1}{8\pi} \int\, \frac{\dd\Omega_3}{4\pi} \, \overline{|\mathcal{M}_{\nu 12 \to l'34}|^2} \,\frac{\tilde{\mathbf{p}}_3^2}{ \left| \left( P_0 - \tilde{E_3}\right)  |\tilde{\mathbf{p}}_3| + \tilde{E_3}\left( |\tilde{\mathbf{p}}_3| - \mathbf{P}\cdot \hat{\tilde{\mathbf{p}}}_3\right) \right| } ~.
\end{eqnarray}
Here $\tilde{p}_3 = |\tilde{\mathbf{p}}_3|$ is determined as the solution of the equation $P_0 - \tilde{E}_3 - \tilde{E}_4 = 0$ and depends on $P_0$, $\mathbf{P}$ and $\cos(\angle(\mathbf{P},\mathbf{p}_3))$; $\hat{\tilde{\mathbf{p}}}_3$ is a unit vector in direction of $\tilde{\mathbf{p}}_3$.
Inserting Eq.~(\ref{Gtilde3}) now into Eq.\ (\ref{d3sdk}) gives, after using the quasiparticle approximation also for the initial nucleons,
\begin{eqnarray} \label{d3sigma1b}
\frac{\dd^3\sigma^{(3)}}{\dd\Omega' \, \dd E'} &=&
   \mbox{ }    \frac{|\mathbf{k}'|}{|\mathbf{k}|}\,g_1 g_2  \,  \mathcal{S}_{34}\mathcal{S}_{12}  \frac{1}{(2\pi)^4 8} \int\, \dd^3x\,\int\,
    \frac{\dd^3p_{1}}{(2\pi)^3 2 E_1} \int\, \frac{
      \dd^3p_{2}}{(2\pi)^3 2 E_2} f_{1}(x,p_{1}) f_2(x,p_2)
      \nonumber \\
    & & \mbox{ }\times \, \int\, \frac{\dd\Omega_3}{4\pi} \, \overline{|\mathcal{M}_{\nu 12 \to l'34}|^2} \,\frac{\tilde{\mathbf{p}}_3^2}{ \left| P_0  |\tilde{\mathbf{p}}_3| - \tilde{E_3}  \mathbf{P}\cdot \hat{\tilde{\mathbf{p}}}_3 \right| } ~.
\end{eqnarray}
where $p_i^0 = \tilde{E}_i$ in $f_i$.
After inserting the spin-degeneracy factors $g_1 = g_2 = 2$, using $\mathcal{S}_{34}\mathcal{S}_{12} = 1/2$ for CC reactions (the initial $pp$ state is not possible for CC reactions) and assuming isospin independence of the CC interaction matrix element and the phase-space distributions, we obtain finally for the triple-differential cross section
\begin{eqnarray} \label{d3sigma1a}
\frac{\dd^3\sigma^{(3)}}{\dd\Omega' \, \dd E'} &=&  \frac{|\mathbf{k}'|}{|\mathbf{k}|} \frac{1}{4(2\pi)^4}  \int \dd^3x\,\int
    \frac{\dd^3p_{1}}{(2\pi)^3 2 E_1} \int \frac{\dd^3p_{2}}{(2\pi)^3 2 E_2}\, f_{1}(x,p_{1}) f_2(x,p_2)  \nonumber \\
    & & \mbox{ }\times \,  \int \frac{\dd\Omega_3}{4\pi}  \,  \overline{|\mathcal{M}_{\nu 12 \to l'34}|^2} \,\frac{\tilde{\mathbf{p}}_3^2}{ \left| P_0  |\tilde{\mathbf{p}}_3| - \tilde{E_3}  \mathbf{P}\cdot \hat{\tilde{\mathbf{p}}}_3 \right| } ~.
\end{eqnarray}
Higher-fold differential cross sections such as Eq.~(\ref{d5sdk}) can be obtained similarly. Numerically the phase-space was evaluated in the cm frame $\mathbf{P} = 0$. The outgoing momenta $p_3$ and $p_4$ were then Lorentz-transformed back to the nuclear rest frame where Pauli-blocking was imposed.

Comparison of Eq.\ (\ref{d3sigma1a}) with Eq.~(\ref{sigma}) allows one to express the nuclear tensor $W^{\mu \lambda}$ for the nuclear systems in terms of that for the $2N$ system,
\begin{equation}    \label{Wmunu}
W^{\mu \lambda} = \mathcal{S}_{12} \int \dd^3x \left[ \int \frac{\dd^3p_1}{(2\pi)^3} \int \frac{\dd^3p_2}{(2\pi)^3} f_{1}(x,p_1) f_2(x,p_2)\,F\, w^{\mu \lambda} \right]~.
\end{equation}
Here $F = (p_{NN} \cdot k)/(p_{NN}^0 E_\nu)$ is the usual flux factor that transforms the cross section from the two-nucleon ($NN$) to the nuclear system, and $w^{\mu \lambda}$ represents the hadronic tensor for the weak interaction with the two-nucleon system.
We neglect the Lorentz transformation from the moving $NN$ system to the stationary nuclear system.
The matrix element $\overline{|\mathcal{M}_{\nu 12 \to l'34}|^2}$ and the hadronic tensor $w^{\mu \lambda}$ are related by
\begin{equation}  \label{M2}
\overline{|\mathcal{M}_{\nu 12 \to l'34}|^2}= L_{\mu \lambda} w^{\mu \lambda}~,
\end{equation}
where $L_{\mu \lambda}$ is the leptonic tensor.
The phase-space distributions $f$ appearing here are solutions of the transport equation (\ref{eq:os-transp.9}) and thus contain the effects of the nuclear potentials as well as of all final-state interactions. Starting from a reliable model for neutrino interaction with the $2N$ system one could calculate
$w^{\mu \lambda}$ consistently in a Fermi-gas model.
\end{widetext}

Eq.\ (\ref{d3sigma1a}) shows nicely [cf.\ Eq.~(\ref{rho})] that the 2p-2h production cross section is $\propto \rho^2$ where $\rho$ is the nuclear density. The fact that both initial nucleons are at the same location $x$ reflects spatial correlations between them. The dependence on the $\rho^2$ also immediately implies that FSI will be very important for this process, both for the inclusive cross section and the semi-inclusive one for knock-out nucleons.

\section{Results}
There are various detailed calculations for the matrix element $\mathcal{M}$ appearing in Eq.~(\ref{d3sigma1a}) presently going on \cite{Amaro:2010sd,Amaro:2011qb,Martini:2009uj,Martini:2011wp,Nieves:2011pp}. These calculations are quite demanding and still have to fight with problems such as gauge invariance, relativity, the inclusion of final state interactions, and limited energy regions of applicability. For example, the authors of Ref.\ \cite{Nieves:2011pp} give as an upper limit of validity of their calculations an energy of 1.0 - 1.2 GeV, whereas the MiniBooNE flux extends well beyond this limit. Since the main purpose of this paper is to look for possible distinctive experimental observables for the presence of 2p-2h processes and to investigate their influence on the energy reconstruction, we now make simplifying assumptions that, as the results will show, still seem to capture most of the relevant physics. The starting point for our approximation is the observation that naturally all neutrino cross sections with nuclei are averaged over the incoming neutrino energy distribution. For the double-differential cross section $\dd^3\sigma / {\dd\Omega' \dd E'}$ at a fixed E' this directly translates into an average over energy transfer and $Q^2$ that will wipe out many details of the matrix element. We, therefore, directly parametrize an average matrix element in terms of the two Mandelstam variables $s$ and $t = -Q^2$ and fit the $s$ dependence to the energy-separated MiniBooNE data in order to fix the overall strength of this contribution at each energy. With this input we then calculate, as a first test, the double-differential cross sections and compare them with the MiniBooNE data. Furthermore, we  investigate the numbers and spectra of knock-out nucleons. Finally, we will discuss the implications for the energy reconstruction.

We have worked with various parametrizations and will give results here only for two. One is a pure phase-space model, in which $\mathcal{M}$ is constant (model I); the constant is fitted to the difference of the energy-separated MiniBooNE data and the calculated QE cross section. In the other model we approximate the hadronic tensor $w$  by the transverse projector $P_{\rm T} = - g_{\mu \nu} - q_\mu q_\nu/Q^2$ (model II), again fitting the overall strength and an $s$ dependence, and then contract it with the lepton tensor. The latter model is meant to mimic the prevalence of  two-body reaction components which are known to be transverse \cite{Boffi:1993gs}. In all these studies RPA effects are not taken into account. Since these tend to lower the cross section in particular at forward angles, our calculations may somewhat underestimate the contributions of 2p-2h effects there. We note, however, that very recent studies show that RPA effects are neglible for neutrino energies larger than 1 GeV and are much smaller than theoretical uncertainties already for $E_\nu > 0.7 \GeV$ \cite{Nieves:2012yz}.

All calculations are done for charged current interactions on a ${}^{12}C$ target, using an axial mass $M_A=1\GeV$ for QE processes.
We evaluate Eq.~(\ref{d3sigma1a})  by picking randomly a pair of nucleons (1,2) at the same location with momenta chosen out of the Fermi sea; all further correlations are assumed to be contained in the matrix element $\mathcal{M}$. Then --- at fixed outgoing lepton energy and angle --- we choose two outgoing nucleons (3,4) that obey energy and momentum conservation, i.e., that are\ weighted according to the integrand of the $\Omega_3$ integration in Eq.~(\ref{d3sigma1a}).  We then follow these two nucleons through and out of the nucleus with all the final-state interactions (elastic and inelastic scattering, charge transfer) included. In doing so we assume that the density and potential of the target nucleus are not significantly disturbed (the so-called frozen approximation). The whole event finally receives a weight given by the cross sections (\ref{d3sigma1a}). This procedure yields both the knock-out cross sections and --- after integration over the momenta of the outgoing nucleons --- also the inclusive double--differential cross section.  We note that the calculations include the FSI of all particles and use fully relativistic kinematics throughout.

\subsection{Inclusive cross sections}

\begin{figure}[hbt]
\includegraphics[width=\columnwidth]{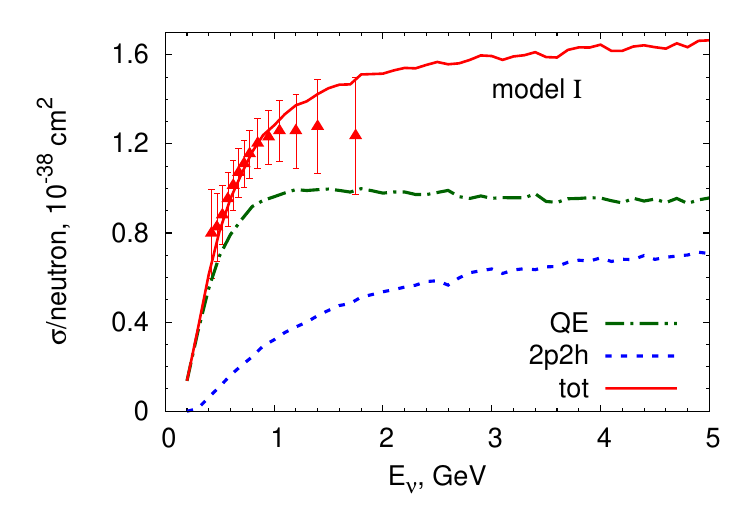}
\caption{(Color online) MiniBooNE data for QE scattering (from Ref.~\protect\cite{AguilarArevalo:2010zc}) together with the result of the GiBUU model for the integrated QE cross section (dash-dotted line), calculated with $M_A = 1 \GeV$.
The dashed line gives the contribution of 2p-2h excitations using a constant matrix element (model I, see text);
the solid line gives the sum of both. All cross sections are per neutron.}
\label{fig:sigmaME4}
\end{figure}

In Fig.\ \ref{fig:sigmaME4} we show for reference again the MiniBooNE data \cite{AguilarArevalo:2010zc} (those with the removed contribution from the absorbed pions) together with the predictions of the GiBUU model for the QE cross section \cite{Leitner:2006ww}. The difference between the data and our QE calculation gives the 2p-2h contribution which is fitted here by model I. It is obvious that this very simple phase-space model is not perfect. It gives a cross section that rises too steeply at energies above about $1\GeV$ (but is still within the experimental errors bars).

\begin{figure}[hbt]
\includegraphics[width=\columnwidth]{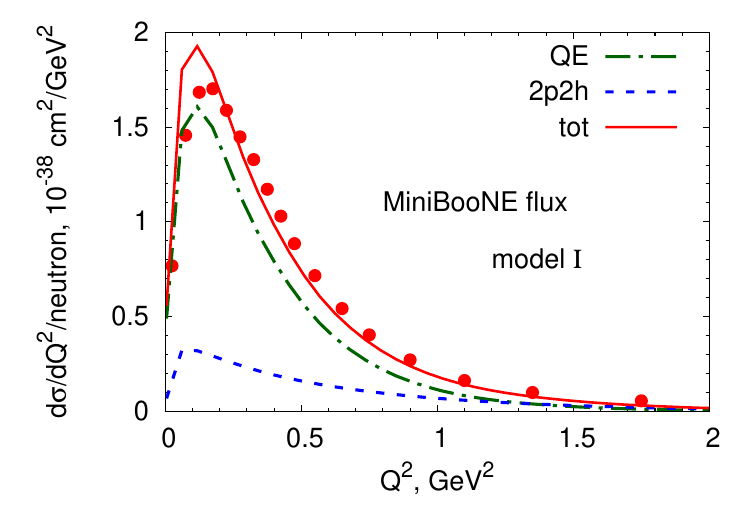}
\caption{(Color online) MiniBooNE data for QE scattering (from Ref.~\protect\cite{AguilarArevalo:2010zc}) together with the result of the GiBUU model for the
$\dd\sigma/\dd Q^2$ cross section per neutron for model I.
All contributions are averaged over the MiniBooNE flux.}
\label{fig:dsigmadQ2ME4}
\end{figure}

Fig.\ \ref{fig:dsigmadQ2ME4} shows the calculated cross section $\dd\sigma/\dd Q^2$ in model I. While the gross structure is quite well reproduced, there is a clear disagreement at lower $Q^2$.
Taking into account the RPA correlations would cure this problem. Indeed, as it was shown in Ref.~\cite{Martini:2011wp}, RPA correlations suppress the true-QE cross section by $20\%$-$25\%$ at $Q^2<0.2 \GeV$ and by a smaller percentage further up to $Q^2<0.4\GeV$. We show this result here only to exhibit the $Q^2$ dependence of the 2p-2h contribution on phase space alone. Phase space alone thus favors an increase of the 2p-2h cross section towards smaller $Q^2$; the sudden fall-off towards $Q^2 = 0$ is caused by Pauli-blocking and phase-space restrictions due to the finite muon mass.

\begin{figure}[hbt]
\includegraphics[width=\columnwidth]{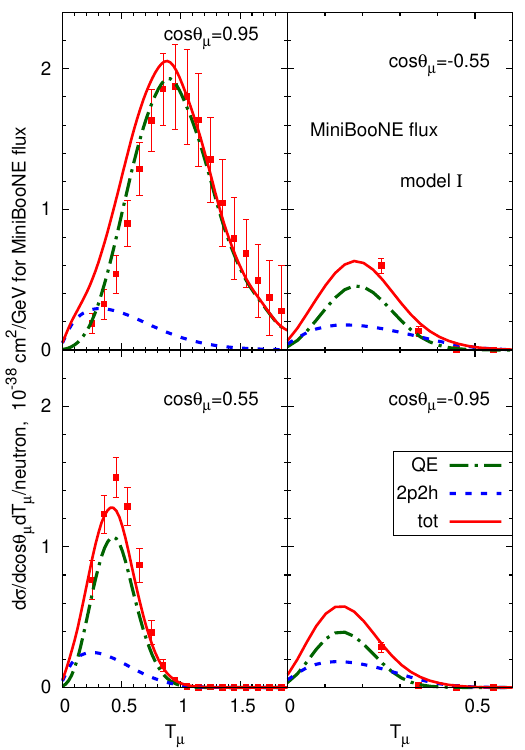}
\caption{(color online) The double-differential cross section $\dd\sigma/(\dd \cos\theta_\mu\,\dd T_\mu)$ per neutron for model I
for four different muon scattering angles versus muon kinetic energy.}
\label{fig:dsigmadcosdE_ME4}
\end{figure}

In order to understand this result we show in Fig.\ \ref{fig:dsigmadcosdE_ME4} the various contributions to the double-differential cross section for four different angles. The cross sections are averaged over the MiniBooNE flux. It is seen, first, that the 'bare' QE contribution describes the forward data very well, but comes out somewhat too low at the higher angles.  It is furthermore seen that the shape and over-all size of the 2p-2h contribution is rather independent of angle (within a factor of 2), amounting to $\approx 0.2 \cdot 10^{-38}\,{\rm cm}^2/\mathrm{GeV}$ in its maximum at $T_\mu \approx 0.2 \GeV$. The overall agreement reached by fitting one number, the size of the squared matrix element in Eq.~(\ref{d3sigma1a}),  is already quite good and certainly better than in much more sophisticated model calculations of the matrix element \cite{Amaro:2010sd}. The main discrepancies appear at very forward angles $\cos \theta = 0.95$, where the cross section is overestimated due to the absence of RPA in our calculations, and at slightly larger angles where the peak cross section is underestimated.

\begin{figure}[hbt]
\includegraphics[width=0.7\columnwidth]{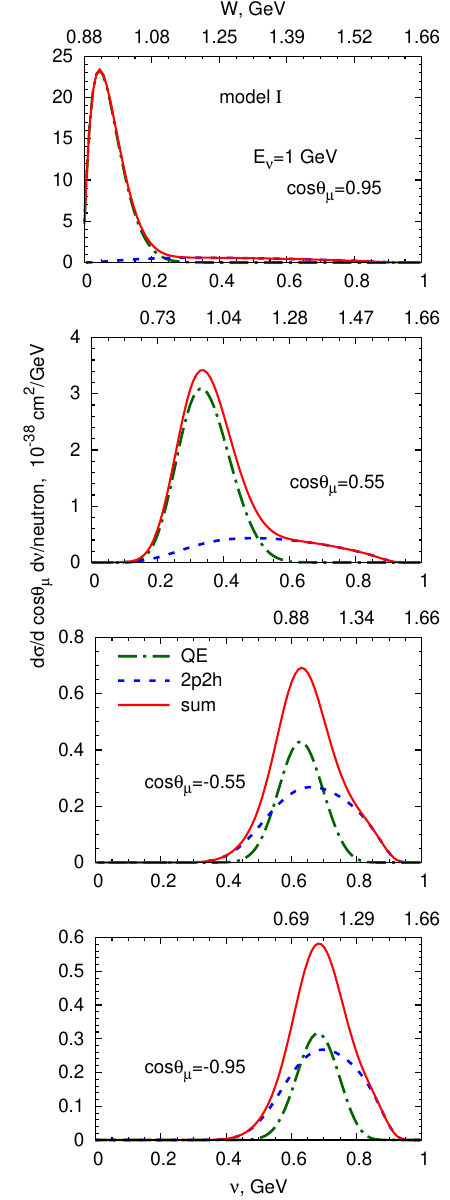}
\caption{(Color online) The cross section $\dd\sigma/(\dd\nu\,\dd\cos\theta_\mu)$ per neutron for model I for four different angles as a function of energy transfer $\nu$. The incoming neutrino energy is fixed at $1 \GeV$.}
\label{fig:dsigmadcosdnu_ME4_fixedenergy}
\end{figure}

Since these results are averaged over the MiniBooNE flux it is instructive to look at the inclusive cross section also for a fixed energy calculation. This is shown in Fig.\ \ref{fig:dsigmadcosdnu_ME4_fixedenergy}, which gives the cross section for the various components at the fixed neutrino energy of $1 \GeV$ as a function of energy transfer $\nu$. For convenience the invariant mass $W$, as it would be determined experimentally from lepton kinematics $W^2=M^2+2M\nu-Q^2$
with $Q^2=2E_\nu(E'-|k'|\cos\theta')$,  is also labeled at the top horizontal axis of each panel. Fig.\ \ref{fig:dsigmadcosdnu_ME4_fixedenergy} shows that the QE contribution drops drastically when going from a forward to a backward angle while the 2p-2h contribution decreases only slightly. This results in a significant rise of the ratio of 2p-2h/QE with angle; at backward angles both contributions become comparable.

\begin{figure}[hbt]
\includegraphics[width=\columnwidth]{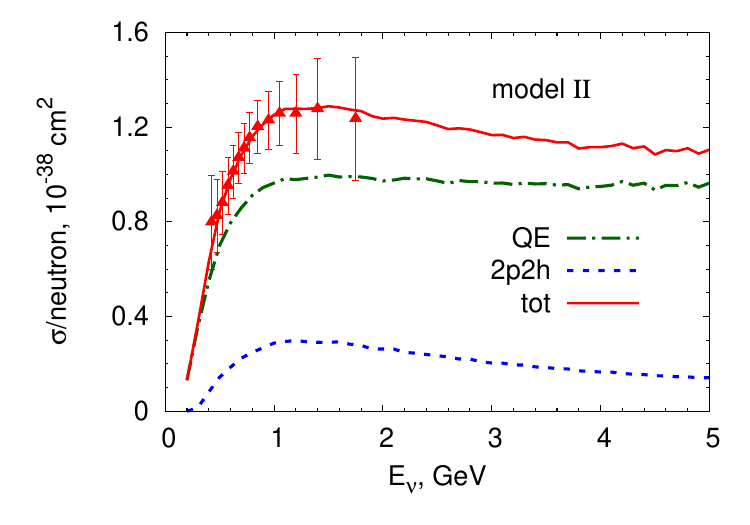}
\caption{(Color online) MiniBooNE data for quasielastic scattering (from \cite{AguilarArevalo:2010zc}) together with the result of GiBUU
for the total cross section per neutron. As Fig.~\ref{fig:sigmaME4}, but now for model II: the hadronic tensor has been assumed to be proportional to the transverse projector and contains an $s$ dependence, see text.}
\label{fig:sigmaME8}
\end{figure}

In order to exhibit the influence of a particular $Q^2$ dependence of the matrix element we now model the hadronic currrent by the transverse projector (model II), again fixing the overall magnitude of the matrix element by fitting the MiniBooNE energy-separated data. In this fit we also allow for an $s$-dependence of the matrix element, which leads to a much better correspondence to the data, as shown in Fig.~\ref{fig:sigmaME8}.

\begin{figure}[hbt]
\includegraphics[width=\columnwidth]{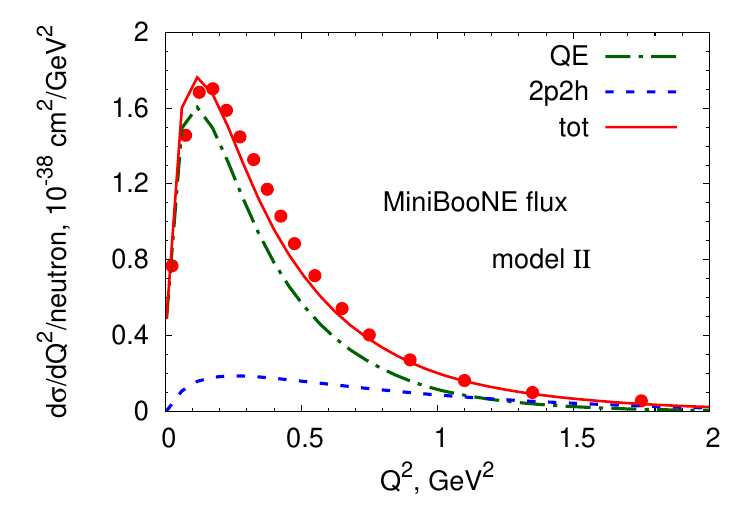}
\caption{(Color online) MiniBooNE data for quasielastic scattering (from Ref.~\cite{AguilarArevalo:2010zc}) together with the result of the GiBUU model for the $\dd\sigma/\dd Q^2$ cross section per neutron for model II (see text).}
\label{fig:dsigmadQ2ME8}
\end{figure}

In Fig.\ \ref{fig:dsigmadQ2ME8} we show again the results for $\dd\sigma/\dd Q^2$.
The agreement is now significantly better, with still a little too much strength at small $Q^2$. However, we note that our calculations do not contain any RPA correlations that tend to lower the cross section at forward angles; the 2p-2h strength should thus probably be even a little larger at forward angles than in our studies here.

\begin{figure*}[hbt]
\includegraphics[width=0.85\textwidth]{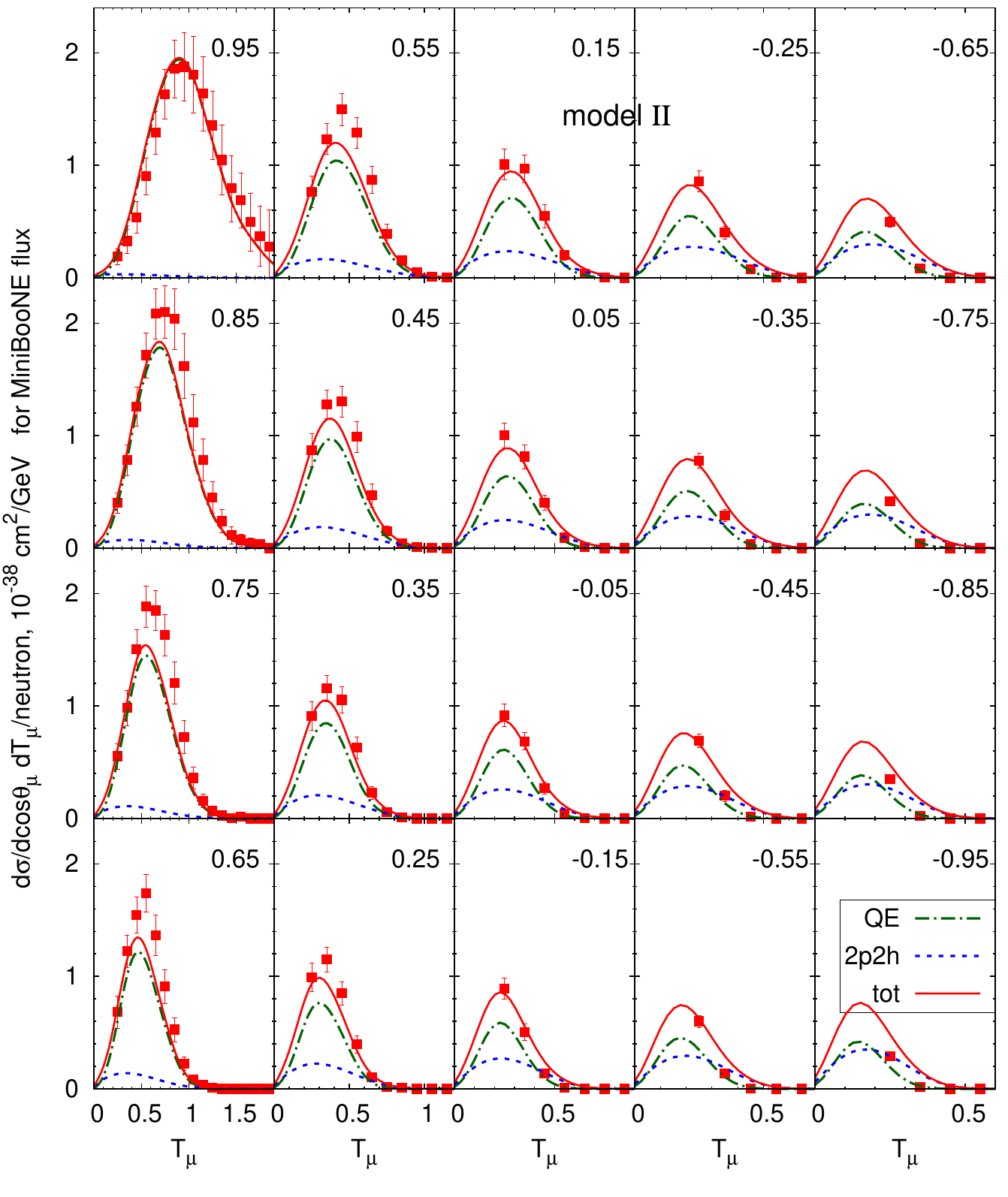}
\caption{(Color online) The double-differential cross section $\dd\sigma/(\dd\cos\theta_\mu\,\dd T_\mu)$ for various muon scattering angles versus
muon kinetic energy for model II. The data are taken from Ref.~\cite{AguilarArevalo:2010zc}.}
\label{fig:dsigmadcosdE_ME8}
\end{figure*}

In Fig.\ \ref{fig:dsigmadcosdE_ME8} we show the double--differential cross sections for all angular bins, as obtained in model II. Immediately evident is the fact that the QE process alone describes the forward--angle cross sections quite well, but becomes significantly too low at backwards angles. Conversely, the 2p-2h contribution is seen to be very small at forward angles and grows with angle so that at backwards angles it is as large as that of the QE process. It is seen that at forward angles the 2p-2h contributions are largest at small $T_\mu$, i.e.,\ at large energy losses below the QE peak, whereas at larger angles the 2p-2h energy loss peaks roughly at the position of the QE peak. Overall, the description reached in our simple model using the MiniBooNE flux is obviously quite good and certainly comparable to that obtained in a microscopic model with flux renormalization (cf.\ Fig.\ 1 in Ref.~\cite{Nieves:2011yp}).

 This agreement with the results of Ref.\ \cite{Nieves:2011yp} as well as the comparison of the results for model I (extreme, phase space only) with those for model II (some physics basis, transverse) shows that the flux-averaging leads to a remarkable insensitivity of the inclusive double-differential cross sections to the input matrix element. In general good agreement is reached with these simple parametrizations of the interaction matrix elements.  It thus seems to be obvious that by a proper parametrization of the $(s,Q^2)$ dependence of the matrix element a nearly perfect agreement for all the observable inclusive cross section data could be reached. While these inclusive double-differential cross sections are 'cleanest' in the sense that they contain only a small model dependence (in the removal of pion production events in which the pion is absorbed in the target) they also seem to be fairly insensitive to details of the hadronic tensor. In particular they do not allow one to separate QE, RPA, and 2p-2h effects in a unique way.

We abstain here from trying to get a perfect fit for the matrix elements and instead now look for observable consequences of the 2p-2h processes, that go beyond just inclusive cross sections with their inherent ambiguities as far as their detailed composition is concerned.

\subsection{Nucleon knock-out}

In this section we thus discuss now properties of knock-out nucleons and their sensitivity to the underlying mechanism (1p-1h vs\ 2p-2h). Hereinafter the calculations contain all relevant processes, i.e., QE, resonance and background pion production, and model II for 2p-2h processes.

\begin{figure}[hbt]
\includegraphics[width=0.7\columnwidth]{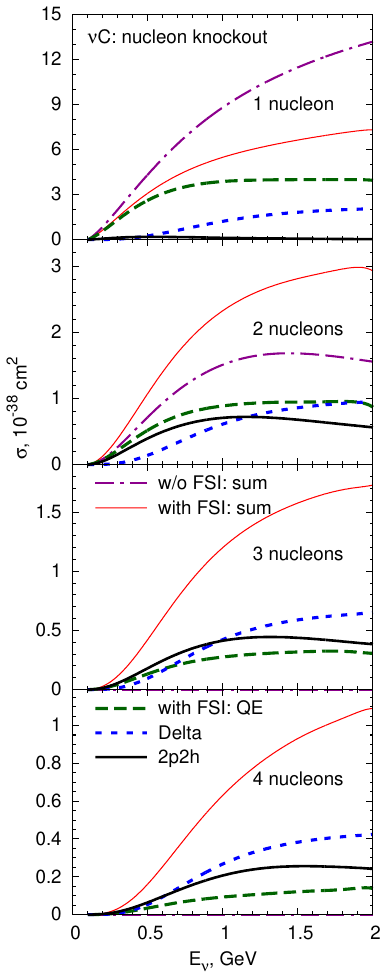}
\caption{(Color online) Cross-section for multi-nucleon knock-out as a function of neutrino energy. Here the hadronic tensor has been assumed to be proportional to the transverse projector (model II).
In each case the total cross section is given by the thin solid line.
The QE contribution is indicated by the long-dashed line. The short-dashed line shows the $\Delta$ contribution, while the 2p-2h contributions is indicated by the thick solid line. All these lines show the results after FSI, while the dot-dashed curves shows the total cross section before FSI.}
\label{fig:sigma(Enu)_multinucleon}
\end{figure}

In Fig.\ \ref{fig:sigma(Enu)_multinucleon} we show the knock-out cross sections for 1, 2, 3, and 4 nucleons before (dot-dashed curves) and  after (thin solid curves) FSI. Various original processes (2p-2h, QE, $\Delta$, higher resonances, 1-pion background) can contribute to the same multiplicity.
For the cross section after FSI some of them (2p-2h, QE, $\Delta$) are explicitly shown.
There is no interference between different contributions to the same multiplicity event because the reactions are all semi-inclusive.

Before FSI, 1-nucleon knock-out comes from all the processes considered in the model. This includes a tiny contribution from 2p-2h processes, which may happen when one of the outgoing nucleons has a too-small kinetic energy and remains bound inside the nucleus. For 2-nucleon knock-out before FSI the contribution is solely from the 2p-2h processes; 3- and 4-nucleon knock-outs are not possible (recall that we do not consider DIS processes here).

After FSI, all the curves are noticeably different from the corresponding ones before FSI. For 1-nucleon knock-out FSI lower the cross section, whereas for all the higher multiplicities they increase it.  In each case the total cross section rises with energy, driven by the QE contribution at the smaller energies $E \leq 0.5 \GeV$ and by pion production above that energy. Only the $\Delta$ contribution to the latter process is explicitly shown in the figure.
After an initial rise the 2p-2h contribution stays rather constant.

For the 2-nucleon knock-out, the QE contribution is even slightly bigger than the 2p-2h one. This is a FSI effect in which after the initial QE process the outgoing nucleon collides with bound nucleons in the nucleus and knocks out one of them.
At the same time the $\Delta$ contribution rises and becomes comparable with both at a neutrino energy of about $1\GeV$.
Two processes are responsible for this rise.
First, the main nonpionic decay channel of $\Delta$ in medium, namely radiationless transition $\Delta N \to NN$, leads to the 2 outgoing nucleons.
Second, both pion and nucleon produced in $\Delta$ decay can collide with bound nucleons in the nucleus and knock out one of them.  This would again give 2 outgoing nucleons. Those again can collide with bound nucleons knocking out the third one, thus leading to an avalanche effect.

The very same processes increase the number of knockout nucleons even more effectively when started by two nucleons
originating from a 2p-2h vertex. They are responsible for the significant decrease of 2p-2h cross section
in the 2-nucleon channel and the increase of that in channels with higher multiplicity.
After FSI, for all 2-, 3- and 4-knockout nucleons the relative 2p-2h contribution amounts to about 20\%.
The relative 2p-2h contribution,  as compared to the QE one, increases with multiplicity, which is also a consequence of the higher starting multiplicity (2 nucleons as compared to 1).

Thus, FSI are strong enough to lead to avalanching that masks the starting event. Indeed all the 3- and 4-nucleon knock-out events are due to FSI.

On the other hand, the topology of the final state of the first, primary interaction (1 versus 2 outgoing nucleons) is sufficiently different for 1p-1h and 2p-2h processes.  Even the avalanchig from FSI does not wash out these differences: 2p-2h processes hardly contribute to the 1-nucleon knock-out.
This implies that the method to absorb many-body effects into a rescaling of the magnetic (single-particle) electromagnetic form factor of the nucleon leads to incorrect final states if used in a standard event generator as suggested by Bodek et al.\ \cite{Bodek:2011ps}.

There are three interesting messages contained in Fig.~\ref{fig:sigma(Enu)_multinucleon}.
First, QE scattering and $\Delta$ contribute significantly not only to 1-, but also to 2-, 3- and 4-nucleon knock-out; 2p-2h processes noticeably contribute to 2-, 3- and 4-nucleon knock-out. Thus, particle multiplicities alone do not distinguish between initial 1p-1h and 2p-2h excitations, which means that  comparison with theory has to be quantitative and not just qualitative.

Second, after  FSI  2p-2h processes contribute less than 30\% to the 2-nucleon knock-out.
Thus, even a perfect theoretical model for 2p-2h interactions, taken on its own, is not sufficient to predict the nucleon output. A realistic description of FSI is essential for any verification of the reaction process through comparison with experimental data.

The third message  is an expected one: since the 1-nucleon channel is practically exclusively due to the initial QE process and pion production, gating on events with one and only one proton (and no pions) can thus help to select true QE scattering. We have shown in Refs.~\cite{Leitner:2010kp,Leitner:2010jv} that even then detector thresholds affect the measured QE cross section significantly. For example, in a tracking detector with a typical detection threshold of proton kinetic energy of $0.2\GeV$  only about one half of the full QE cross section is actually measured; the rest has to be reconstructed with the help of an event generator.

\begin{figure}[hbt]
\includegraphics[width=0.7\columnwidth]{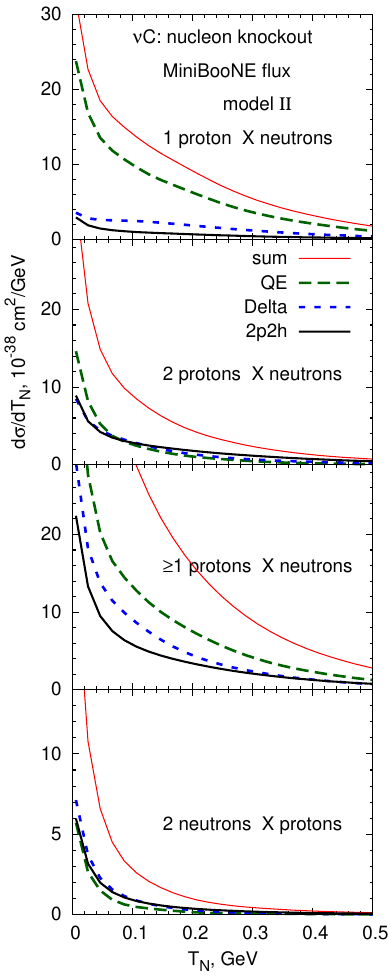}
\caption{(color online) Kinetic energy spectra for knocked-out protons (three upper panels) and neutrons (bottom panel) for events with any number of mesons in the final state for the MiniBooNE flux.}
\label{fig:n-knockout}
\end{figure}

Finally, in Fig.\ \ref{fig:n-knockout} we show the kinetic energy spectra of knock-out nucleons for events with  1 proton and any number of neutrons,  2 protons and any number of neutrons, at least 1 proton and any number of neutrons, and 2 neutrons and any number of protons in the final state for the MiniBooNE flux. First of all, the plots show that the importance of 2p-2h knock-outs relative to others increases dramatically with the number of ejected protons. While it is insignificant for the ``1 proton $X$ neutrons'' events it is of comparable size for the higher-multiplicity events. Second, there is a pileup of cross section at small kinetic energies $T_N$ (below about $0.1\GeV$) due to FSI; the 2p-2h contributions alone exhibit a similar behavior, though with a slightly smaller rise at small $T_N$. The 2p-2h contributions increase the cross sections at small kinetic energies significantly, by a factor 2 - 3.

\begin{figure}[hbt]
\includegraphics[width=0.7\columnwidth]{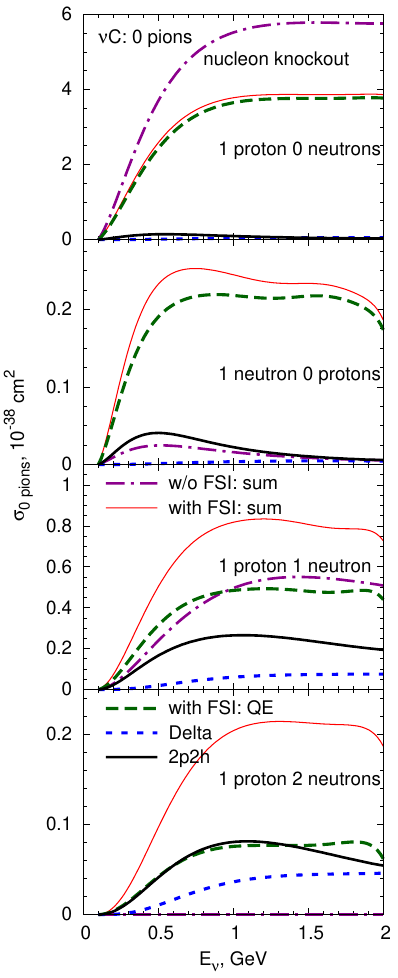}
\caption{(Color online) Cross section for nucleon knock-out as a function of neutrino energy for events with 0 pions in the final state.}
\label{fig:sigmaEnu_0pions_multinucleon}
\end{figure}

As already discussed, of primary interest for the modern experiments are the QE-like events, which by experimental definition for \v{C}erenkov detectors
comprise all events with no pions in the final state. In Fig.~\ref{fig:sigmaEnu_0pions_multinucleon} we show the nucleon knock-out cross sections for these events. Similar to the previous discussion, we compare the calculations before and after FSI.

For the final state with 1 proton and 0 neutrons the by-far dominant contribution comes from true QE scattering. A tiny contribution from the 2p-2h processes (as discussed above, this is possible if one of the outgoing nucleons has a very low kinetic energy) is negligible. FSI decrease the cross section; this is because a higher-energy nucleon may rescatter and knock out more nucleons from the nucleus. For the final state with 0 protons and 1 neutron the only contribution before FSI comes from the 2p-2h processes, if one of the nucleons remains bound. FSI of outgoing protons coming from the intial QE process increase this cross section by a factor of 5. The final state with 1 proton and 1 nucleon before FSI is again solely of 2p-2h origin. FSI noticeably change the weight of various contributions: the output from 2p-2h is reduced, while QE process contributes significantly. A final state with 1 proton and 2 neutrino is only possible after FSI.
Here one also sees a mixture of events originating from QE, $\Delta$, and 2p-2h.

Thus, the only clear signature of true QE scattering are the events with 0 pions, 1 proton, and 0 neutrons in the final state. The purity of such an
event sample would be very high, as the nearly negligible difference between the dashed and thin solid curves in Fig.\ \ref{fig:sigmaEnu_0pions_multinucleon} shows; however, the selection efficiency is only at the level of $\approx 70\%$ as shown by the  difference between the dashed and dash-dotted curves.
Unfortunately, most detectors are insensitive to neutrons, thus making the experimental selection of true QE events practically impossible
and forcing one to use Monte Carlo simulations for reconstruction. Gating on events with 1 proton, 0 pions and any number of (undetected) neutrons
leads to a an event sample with a purity at the level of $\approx 90\%$.

\subsection{Energy reconstruction}

\begin{figure}[hbt]
\includegraphics[width=0.9\columnwidth]{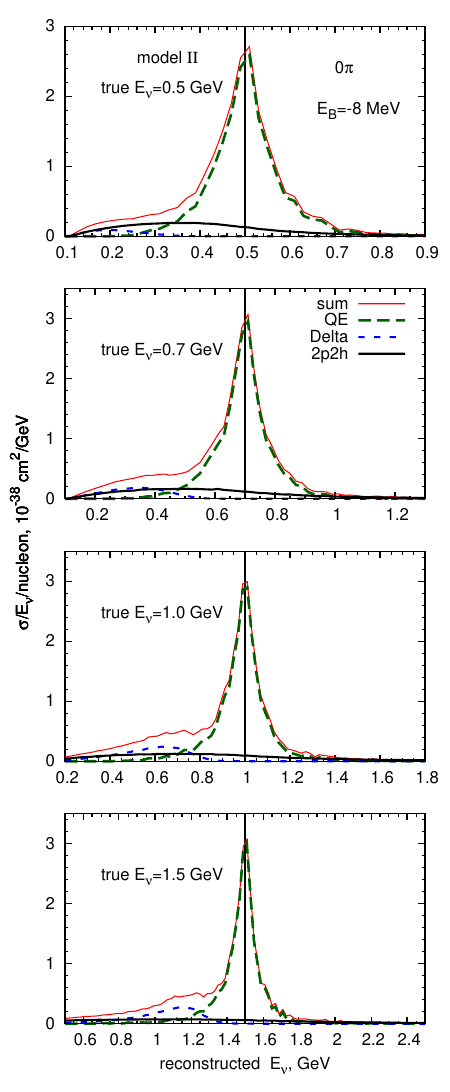}
\caption{(Color online) Distribution of the reconstructed neutrino energy  for QE-like processes as defined by the MiniBooNE experiment (0 pions and any number of nucleons in the final states) for the true energy of the incoming neutrino of $0.5, \ 0.7, \ 1.0$, and $1.5 \GeV$.
}
\label{fig:enu-reconstruction-fixedEnu-nopi-ME8}
\end{figure}

Many neutrino experiments, including MiniBooNE, use QE scattering as a tool to determine the incoming neutrino energy. This is usually done by using quasifree kinematics for the QE process on a neutron at rest. It is obvious that Fermi motion is thus neglected from the outset and can be expected to lead to a broadening of the reconstructed energy around the true neutrino energy. Any binding effects in these analyses are taken into account by assuming a constant energy shift. In Ref.~\cite{Leitner:2010kp} we have shown that already the entanglement of QE scattering and pion production leads to a significant downward shift of the reconstructed energy. In this section we now analyze the influence of the initial 2p-2h excitations on the energy reconstruction. Based on the implementation of 2p-2h effects as described above we have analyzed the energy reconstruction on an event-by-event basis. The calculations contain all relevant reaction mechanisms, in this case QE scattering and pion production, and include final-state interactions.

The relevant formula used for the energy reconstruction is based on the assumption of QE scattering on a nucleon at rest \cite{AguilarArevalo:2008yp},
\begin{equation}   \label{Ereconstr}
E_\nu = \frac{2(M_n - E_B)E' - (E_B^2 - 2M_n E_B + m_\mu^2 + \Delta M^2)}{2\left[M_n - E_B - E' + |\vec{k}'| \cos \theta_\mu\right]} ~.
\end{equation}
Here $E_B > 0$ is a constant binding energy, $M_n$ the mass of the neutron, and $\Delta M^2 = M_n^2 - M_p^2$. $E'$, $|\vec{k}'|$, and $\theta_\mu$ are the energy, momentum, and angle of the outgoing muon.

There are two features of this expression that affect the analysis of a 2p-2h event using Eq.~(\ref{Ereconstr}), which is correct only for true QE scattering, i.e.,\ a 1p-1h event. First, in the last section we have shown that, at forward angles, where the cross section is largest, the 2p-2h contributions are largest at small $T_\mu$, below the QE peak. When analyzing such events with the help of the one-body expression (\ref{Ereconstr}) this leads to a lower reconstructed neutrino energy than the true one. Second, while QE scattering is strongly forward peaked, the 2p-2h events are fairly flat (within a factor of 2) in lepton angle (see Figs.\ \ref{fig:dsigmadcosdE_ME4} and \ref{fig:dsigmadcosdE_ME8}). The relatively strong  yield at backward angles will lead to a larger neutrino energy, in particular for intermediate muon energies. Since both effects are present we expect a fairly flat behavior of the 2p-2h contribution to the energy reconstruction.

In Fig.~\ref{fig:enu-reconstruction-fixedEnu-nopi-ME8} we plot the distribution of the reconstructed neutrino energy obtained using the MiniBooNE reconstruction method with  $E_B=8 \MeV$ (which is a typical binding energy in the GiBUU code, as opposed to the value $34 \MeV$ used by  MiniBooNE) for the fixed true neutrino energies $E_\nu^\mathrm{true}=0.5, \ 0.7, \ 1.0$, and $1.5\GeV$. The main contribution to all QE-like events is given by the true QE events, which shows a prominent peak around the true energy. The peak is approximately symmetric and has a width of about $0.1\GeV$: this broadening is caused by Fermi motion. For $\Delta-$induced events the distribution is not symmetric, with a broad peak at lower energies. For a more detailed discussion of these effects see Refs.~\cite{Leitner:2010jv,Leitner:2010kp}.

For all energies the broad 2p-2h contribution is peaked around 1/2 of the true energy. For the lower neutrino energies of $0.5$ and $0.7\GeV$ most of the total distortion is caused by 2p-2h processes because at this low energy there is only little $\Delta$ excitation. For the true neutrino energies of 1.0 and $1.5\GeV$, on the other hand, the contributions of 2p-2h events are comparably small because here $\Delta$ excitation plays a major role and because their strength is distributed over the whole energy range from $0.2$ to $1.8\GeV$, with a flat maximum again around 1/2 of the true energy. Both the 2p-2h effects as well as the $\Delta$ excitation lead to a shift of the reconstructed energy towards smaller values, or, vice versa, for a given reconstructed energy the true energy always lies higher than the reconstructed one. The effect is most pronounced at low true energies and becomes smaller at higher energies. Both of these results agree with the recent analyses by  Martini et al.\ \cite{Martini:2012fa} and with the very recent results of Nieves et al.\ \cite{Nieves:2012yz}. Both of these publications do contain some treatment of RPA correlations which are most prevalent at lower energies $E_\nu < 0.7 \GeV$ and at forward angles. On the other hand, the work \cite{Nieves:2012yz} does not contain any pion (or $\Delta$) degrees of freedom and thus misses part of the lower hump in Fig.\ \ref{fig:enu-reconstruction-fixedEnu-nopi-ME8}.  This is of no concern as soon as calculations are compared to the cross section with subtracted  pion-induced QE-like background (as done by  MiniBooNE), but is essential for any extraction of oscillation parameters.

\begin{figure}[hbt]
\includegraphics[width=\columnwidth]{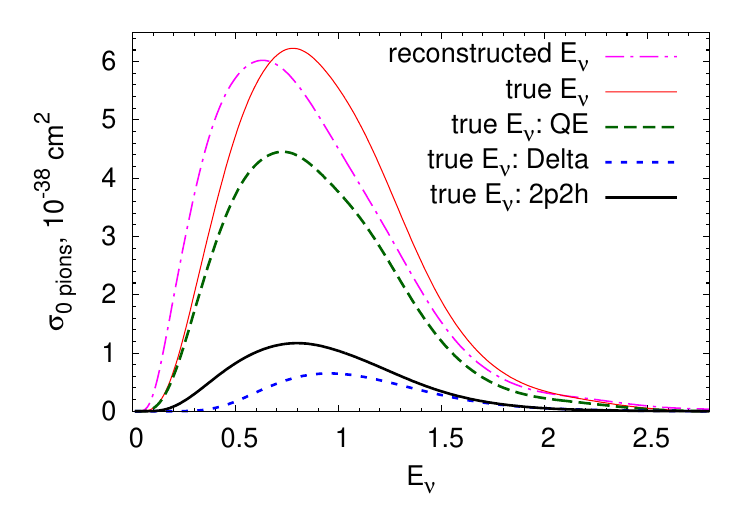}
\\
\includegraphics[width=\columnwidth]{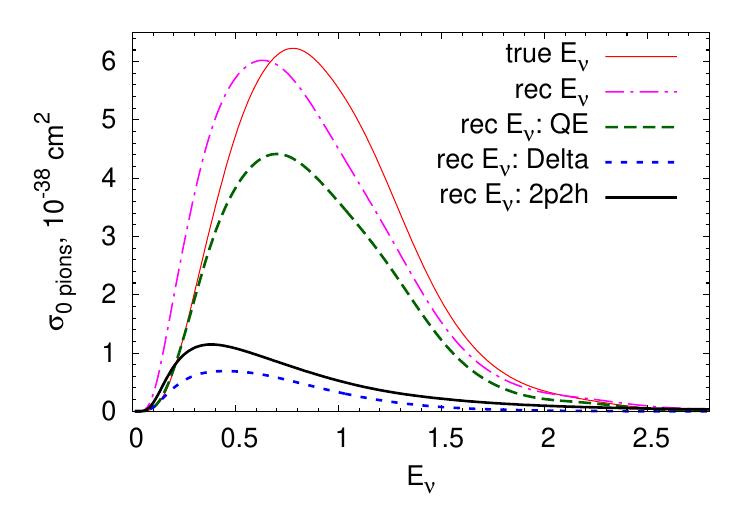}
\caption{(Color online) Event distribution  for QE-like processes as defined by the MiniBooNE experiment versus the true (upper picture) and reconstructed (lower picture) neutrino energy. The thin solid and dot-dashed curves repeat the distributions for true and reconstructed energy in both figures.}
\label{fig:enu-reconstruction-MB-ME8}
\end{figure}
This is also seen in Fig.\ \ref{fig:enu-reconstruction-MB-ME8} which shows the event distribution
(also called flux-folded  cross section) for QE-like processes that would be observed by the MiniBooNE experiment.
Only part of these zero-pion events  are genuine QE events. The rest can originate from initial $\Delta$ production, higher resonance production (not shown in the figure), pion-background events (not shown in the figure), or 2p-2h processes. Since in the numerical simulation we know the true neutrino energy and calculate the reconstructed energy as done by MiniBooNE, we can predict the event distribution versus both of them. The event distribution versus the true energy is shown by the curve labeled 'true $E_\nu$'. Comparing it to the one versus the reconstructed energy one sees that there is a systematic distortion in energy reconstruction: the peak of the event distribution is shifted now by about 100 MeV to \emph{lower} energies with this shift becoming smaller for energies above $1.3\GeV$. Qualitatively this same behavior was already observed and discussed for the K2K and the MiniBooNE flux in Ref.~\cite{Leitner:2010kp} as a consequence of pionic excitations alone. Here, now also the 2p-2h contribution is included in this analysis; it leads to the noticeable downward shift of the reconstructed energy curve on the low-energy side of the maximum where the pion production is still small.\footnote{We note that there is a slight inconsistency in our analysis because we have fitted the matrix element to the cross section vs.\ the reconstructed energy in Figs.\ \ref{fig:sigmaME4} and \ref{fig:sigmaME8}. The error connected with this is still well within the general uncertainties in the figures mentioned.} This distortion is important since it directly distorts the oscillation pattern and thus affects the extraction of the CP-violating phase $\Delta_{\rm CP}$.

So far we have concentrated on a discussion of 2p-2h effects in the MiniBooNE experiment, i.e.,\ a \v{C}erenkov counter experiment. For completeness we now discuss also the K2K experiment which uses a tracking detector, which is sensitive to protons but not neutrons.\footnote{The presently running T2K experiment employs both techniques in its (different) near- and far-side detectors.} In this experiment an event is identified as QE-like if no pions and exactly 1 proton is found in the final state. Evidently, the proton can be accompanied by any number of unobserved neutrons. As we have shown in Ref.~\cite{Leitner:2010kp}, in such a detector there is nearly no distortion from $\Delta$ excitation, but the total QE rate is underestimated (because of charge exchange in the FSI) and has to be reconstructed by means of event generators.

\begin{figure}[hbt]
\includegraphics[width=0.9\columnwidth]{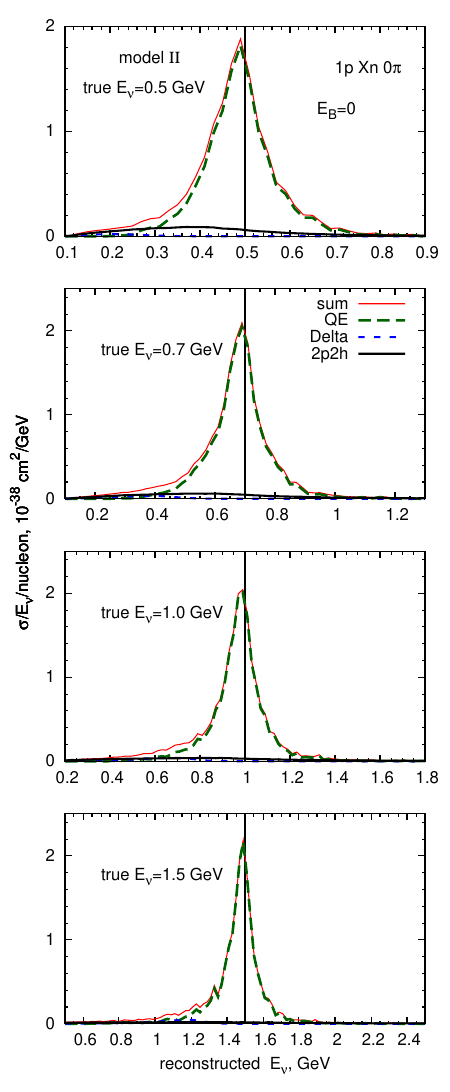}
\caption{(Color online) Distribution of the reconstructed neutrino energy for QE-like processes as defined by the K2K experiment (0 pions, 1 proton, and any number of neutrons in the final states) versus the reconstructed neutrino energy for the true energy of the incoming neutrino of $0.5, \ 0.7, \ 1.0$, and $1.5 \GeV$.}
\label{fig:enu-reconstruction-fixedEnu-p-Xn-nopi-ME8}
\end{figure}

In this experiment the neutrino energy is reconstructed according to Eq.~(\ref{Ereconstr}), but neglecting the binding energy, i.e.,\ with $E_B=0$. Our results for the cross sections versus the reconstructed energy are shown in Fig.~\ref{fig:enu-reconstruction-fixedEnu-p-Xn-nopi-ME8}. One can easily conclude, that the experimental possibility to detect a proton in the final state and to select events with one proton only significantly reduce the contribution from the initial $\Delta$ resonance production, while the distortion from the 2p-2h processes (downward shift) remains on the same level as for the MiniBooNE \v{C}erenkov detector.

\section{Summary}
In summary, understanding the reaction mechanism of neutrinos with nuclei is mandatory for a reliable determination of the neutrino energy. The large excess of QE-like events observed by MiniBooNE over those calculated by commonly used event generators with the world-average value for the axial mass of around 1.0 GeV has therefore led to a multitude of partly contradictory models that all aim to describe the same data, starting from quite different initial 1p-1h or 2p-2h interaction mechanisms. Since broad energy-band neutrino experiments necessarily contain an averaging over quite different reaction mechanisms, the inclusive data alone do not allow for an unambiguous experimental verification of any particular model.

We have, therefore, investigated the effects of 2p-2h interactions by extending the GiBUU event generator, which contains all the relevant reaction mechanisms, on both inclusive and more exclusive observables. Because of the inherent kinematical averaging in any broad-band neutrino experiment  we have approximated the matrix element for neutrino interactions with two nucleons by an ad-hoc average ansatz, either using just a simple phase-space model or a model in which the hadronic tensor is modeled by the transverse projector for the $W$ bosons. In both cases quite reasonable descriptions of the inclusive double-differential cross section depending on muon observables only are obtained. We interpret this result as an indication that inclusive neutrino cross sections in broad energy band experiments are not sensitive to details of the $\nu NN$ interaction as long as average properties of the interaction are described well enough.

Within this same model we have then calculated the numbers and spectra of knock-out nucleons. We find that indeed 2p-2h primary interactions increase the cross sections for multi-nucleon knock-out. Both the numbers and the spectra of knock-out nucleons are changed so that detailed measurements should be able to pin down the strength of the 2p-2h processes. However, even in the presence of more complicated initial interactions the one-proton knockout channel is only slightly affected by 2p-2h contributions. Experiments that can actually restrict the number of knock-out nucleons to one and only one proton should, therefore, be able to provide a clean
(with purity at the level of $90\%$) sample of QE events. We have shown earlier that such events also are not significantly contaminated by $\Delta$ excitations \cite{Leitner:2010kp}; they do miss, however, part of the QE cross section because of charge changing FSI on the outgoing nucleons.
The selection efficiency is expected to be at the level of $70\%$.

Finally, we have shown that the energy reconstruction is affected by the presence of 2p-2h interactions mainly at the lower neutrino energies, while the $\Delta$ excitations become more essential at the higher energies. 2p-2h interactions lead -- as the $\Delta$ excitations do -- to a downward shift of the reconstructed energy. Their contribution to the reconstructed energy distribution is most pronounced at the lower energies and fairly flat over a wide reconstructed energy range for the higher true energies. This result agrees with the recent analyses by Martini et al.\cite{Martini:2012fa} and by Nieves et al.~\cite{Nieves:2012yz} which are based on a microscopical model of the matrix element; we take this as a further indication that our model contains the most relevant physics for this reconstruction.

While we have performed such analyses only for CC processes we note that the formalism can also be applied to NC reactions. For the latter a good description of the final state particles is absolutely essential for the energy and $Q^2$ determination.

Since events involving pion production and $\Delta$ excitations are closely entangled with true QE scattering and 2p-2h processes, any comparison with data must describe at least these reaction mechanisms equally well. In all the calculations of the 2p-2h process reported in this paper we have always assumed that the outgoing two particles from the first, initial interaction are two nucleons. In principle there could, however, also be events such as $\nu + N_1 + N_2 \to l' + N_3 + \Delta$, or even events with two $\Delta$'s in the outgoing state. Such processes would contribute to pion production and enhance the pion yield.  A solution to the puzzle of high pion production cross sections mentioned in the introduction might thus come from the inclusion of 2p-2h events with one or two $\Delta$ in the final state. We speculate that this might solve the MiniBooNE ``pion problem`` shortly mentioned in the Introduction.

\acknowledgments
We are grateful for many helpful comments by and discussions with L.\ Alvarez-Ruso. We also gratefully
acknowledge helpful discussions with M.\ Kaskulov and A.\ Larionov.
This work has been supported by DFG and the Helmholtz International Center for FAIR.

\bibliographystyle{apsrev4-1}
\bibliography{nuclear}

\begin{thebibliography}{59}%
\makeatletter
\providecommand \@ifxundefined [1]{%
 \@ifx{#1\undefined}
}%
\providecommand \@ifnum [1]{%
 \ifnum #1\expandafter \@firstoftwo
 \else \expandafter \@secondoftwo
 \fi
}%
\providecommand \@ifx [1]{%
 \ifx #1\expandafter \@firstoftwo
 \else \expandafter \@secondoftwo
 \fi
}%
\providecommand \natexlab [1]{#1}%
\providecommand \enquote  [1]{``#1''}%
\providecommand \bibnamefont  [1]{#1}%
\providecommand \bibfnamefont [1]{#1}%
\providecommand \citenamefont [1]{#1}%
\providecommand \href@noop [0]{\@secondoftwo}%
\providecommand \href [0]{\begingroup \@sanitize@url \@href}%
\providecommand \@href[1]{\@@startlink{#1}\@@href}%
\providecommand \@@href[1]{\endgroup#1\@@endlink}%
\providecommand \@sanitize@url [0]{\catcode `\\12\catcode `\$12\catcode
  `\&12\catcode `\#12\catcode `\^12\catcode `\_12\catcode `\%12\relax}%
\providecommand \@@startlink[1]{}%
\providecommand \@@endlink[0]{}%
\providecommand \url  [0]{\begingroup\@sanitize@url \@url }%
\providecommand \@url [1]{\endgroup\@href {#1}{\urlprefix }}%
\providecommand \urlprefix  [0]{URL }%
\providecommand \Eprint [0]{\href }%
\providecommand \doibase [0]{http://dx.doi.org/}%
\providecommand \selectlanguage [0]{\@gobble}%
\providecommand \bibinfo  [0]{\@secondoftwo}%
\providecommand \bibfield  [0]{\@secondoftwo}%
\providecommand \translation [1]{[#1]}%
\providecommand \BibitemOpen [0]{}%
\providecommand \bibitemStop [0]{}%
\providecommand \bibitemNoStop [0]{.\EOS\space}%
\providecommand \EOS [0]{\spacefactor3000\relax}%
\providecommand \BibitemShut  [1]{\csname bibitem#1\endcsname}%
\let\auto@bib@innerbib\@empty
\bibitem [{\citenamefont {Arrington}\ \emph {et~al.}(2007)\citenamefont
  {Arrington}, \citenamefont {Roberts},\ and\ \citenamefont
  {Zanotti}}]{Arrington:2006zm}%
  \BibitemOpen
  \bibfield  {author} {\bibinfo {author} {\bibfnamefont {J.}~\bibnamefont
  {Arrington}}, \bibinfo {author} {\bibfnamefont {C.}~\bibnamefont {Roberts}},
  \ and\ \bibinfo {author} {\bibfnamefont {J.}~\bibnamefont {Zanotti}},\ }\href
  {\doibase 10.1088/0954-3899/34/7/S03} {\bibfield  {journal} {\bibinfo
  {journal} {J.Phys.G}\ }\textbf {\bibinfo {volume} {G34}},\ \bibinfo {pages}
  {S23} (\bibinfo {year} {2007})},\ \Eprint
  {http://arxiv.org/abs/nucl-th/0611050} {arXiv:nucl-th/0611050 [nucl-th]}
  \BibitemShut {NoStop}%
\bibitem [{\citenamefont {Bernard}\ \emph {et~al.}(2002)\citenamefont
  {Bernard}, \citenamefont {Elouadrhiri},\ and\ \citenamefont
  {Meissner}}]{Bernard:2001rs}%
  \BibitemOpen
  \bibfield  {author} {\bibinfo {author} {\bibfnamefont {V.}~\bibnamefont
  {Bernard}}, \bibinfo {author} {\bibfnamefont {L.}~\bibnamefont
  {Elouadrhiri}}, \ and\ \bibinfo {author} {\bibfnamefont {U.}~\bibnamefont
  {Meissner}},\ }\href {\doibase 10.1088/0954-3899/28/1/201} {\bibfield
  {journal} {\bibinfo  {journal} {J.Phys.G}\ }\textbf {\bibinfo {volume}
  {G28}},\ \bibinfo {pages} {R1} (\bibinfo {year} {2002})},\ \Eprint
  {http://arxiv.org/abs/hep-ph/0107088} {arXiv:hep-ph/0107088 [hep-ph]}
  \BibitemShut {NoStop}%
\bibitem [{\citenamefont {Aguilar-Arevalo}\ \emph {et~al.}(2008)\citenamefont
  {Aguilar-Arevalo} \emph {et~al.}}]{:2007ru}%
  \BibitemOpen
  \bibfield  {author} {\bibinfo {author} {\bibfnamefont {A.}~\bibnamefont
  {Aguilar-Arevalo}} \emph {et~al.} (\bibinfo {collaboration} {MiniBooNE
  Collaboration}),\ }\href {\doibase 10.1103/PhysRevLett.100.032301} {\bibfield
   {journal} {\bibinfo  {journal} {Phys.Rev.Lett.}\ }\textbf {\bibinfo {volume}
  {100}},\ \bibinfo {pages} {032301} (\bibinfo {year} {2008})},\ \Eprint
  {http://arxiv.org/abs/0706.0926} {arXiv:0706.0926 [hep-ex]} \BibitemShut
  {NoStop}%
\bibitem [{\citenamefont {Aguilar-Arevalo}\ \emph
  {et~al.}(2010{\natexlab{a}})\citenamefont {Aguilar-Arevalo} \emph
  {et~al.}}]{AguilarArevalo:2010zc}%
  \BibitemOpen
  \bibfield  {author} {\bibinfo {author} {\bibfnamefont {A.~A.}\ \bibnamefont
  {Aguilar-Arevalo}} \emph {et~al.} (\bibinfo {collaboration} {MiniBooNE}),\
  }\href {\doibase 10.1103/PhysRevD.81.092005} {\bibfield  {journal} {\bibinfo
  {journal} {Phys. Rev.}\ }\textbf {\bibinfo {volume} {D81}},\ \bibinfo {pages}
  {092005} (\bibinfo {year} {2010}{\natexlab{a}})},\ \Eprint
  {http://arxiv.org/abs/1002.2680} {arXiv:1002.2680 [hep-ex]} \BibitemShut
  {NoStop}%
\bibitem [{\citenamefont {Aguilar-Arevalo}\ \emph
  {et~al.}(2010{\natexlab{b}})\citenamefont {Aguilar-Arevalo} \emph
  {et~al.}}]{AguilarArevalo:2010cx}%
  \BibitemOpen
  \bibfield  {author} {\bibinfo {author} {\bibfnamefont {A.}~\bibnamefont
  {Aguilar-Arevalo}} \emph {et~al.} (\bibinfo {collaboration} {MiniBooNE
  Collaboration}),\ }\href {\doibase 10.1103/PhysRevD.82.092005} {\bibfield
  {journal} {\bibinfo  {journal} {Phys.Rev.D}\ }\textbf {\bibinfo {volume}
  {82}},\ \bibinfo {pages} {092005} (\bibinfo {year} {2010}{\natexlab{b}})},\
  \Eprint {http://arxiv.org/abs/1007.4730} {arXiv:1007.4730 [hep-ex]}
  \BibitemShut {NoStop}%
\bibitem [{\citenamefont {Gran}\ \emph {et~al.}(2006)\citenamefont {Gran} \emph
  {et~al.}}]{Gran:2006jn}%
  \BibitemOpen
  \bibfield  {author} {\bibinfo {author} {\bibfnamefont {R.}~\bibnamefont
  {Gran}} \emph {et~al.} (\bibinfo {collaboration} {K2K}),\ }\href {\doibase
  10.1103/PhysRevD.74.052002} {\bibfield  {journal} {\bibinfo  {journal} {Phys.
  Rev.}\ }\textbf {\bibinfo {volume} {D74}},\ \bibinfo {pages} {052002}
  (\bibinfo {year} {2006})},\ \Eprint {http://arxiv.org/abs/hep-ex/0603034}
  {arXiv:hep-ex/0603034} \BibitemShut {NoStop}%
\bibitem [{\citenamefont {Lyubushkin}\ \emph {et~al.}(2009)\citenamefont
  {Lyubushkin} \emph {et~al.}}]{Lyubushkin:2008pe}%
  \BibitemOpen
  \bibfield  {author} {\bibinfo {author} {\bibfnamefont {V.}~\bibnamefont
  {Lyubushkin}} \emph {et~al.} (\bibinfo {collaboration} {NOMAD}),\ }\href
  {\doibase 10.1140/epjc/s10052-009-1113-0} {\bibfield  {journal} {\bibinfo
  {journal} {Eur. Phys. J.}\ }\textbf {\bibinfo {volume} {C63}},\ \bibinfo
  {pages} {355} (\bibinfo {year} {2009})},\ \Eprint
  {http://arxiv.org/abs/0812.4543} {arXiv:0812.4543 [hep-ex]} \BibitemShut
  {NoStop}%
\bibitem [{\citenamefont {Benhar}\ \emph {et~al.}(2010)\citenamefont {Benhar},
  \citenamefont {Coletti},\ and\ \citenamefont {Meloni}}]{Benhar:2010nx}%
  \BibitemOpen
  \bibfield  {author} {\bibinfo {author} {\bibfnamefont {O.}~\bibnamefont
  {Benhar}}, \bibinfo {author} {\bibfnamefont {P.}~\bibnamefont {Coletti}}, \
  and\ \bibinfo {author} {\bibfnamefont {D.}~\bibnamefont {Meloni}},\ }\href
  {\doibase 10.1103/PhysRevLett.105.132301} {\bibfield  {journal} {\bibinfo
  {journal} {Phys.Rev.Lett.}\ }\textbf {\bibinfo {volume} {105}},\ \bibinfo
  {pages} {132301} (\bibinfo {year} {2010})},\ \Eprint
  {http://arxiv.org/abs/1006.4783} {arXiv:1006.4783 [nucl-th]} \BibitemShut
  {NoStop}%
\bibitem [{\citenamefont {Benhar}(2011)}]{Benhar:2011ef}%
  \BibitemOpen
  \bibfield  {author} {\bibinfo {author} {\bibfnamefont {O.}~\bibnamefont
  {Benhar}},\ }\href@noop {} {\  (\bibinfo {year} {2011})},\ \Eprint
  {http://arxiv.org/abs/1110.1835} {arXiv:1110.1835 [hep-ph]} \BibitemShut
  {NoStop}%
\bibitem [{\citenamefont {Leitner}\ and\ \citenamefont
  {Mosel}(2010{\natexlab{a}})}]{Leitner:2010kp}%
  \BibitemOpen
  \bibfield  {author} {\bibinfo {author} {\bibfnamefont {T.}~\bibnamefont
  {Leitner}}\ and\ \bibinfo {author} {\bibfnamefont {U.}~\bibnamefont
  {Mosel}},\ }\href {\doibase 10.1103/PhysRevC.81.064614} {\bibfield  {journal}
  {\bibinfo  {journal} {Phys.Rev.C}\ }\textbf {\bibinfo {volume} {81}},\
  \bibinfo {pages} {064614} (\bibinfo {year} {2010}{\natexlab{a}})},\ \Eprint
  {http://arxiv.org/abs/1004.4433} {arXiv:1004.4433 [nucl-th]} \BibitemShut
  {NoStop}%
\bibitem [{\citenamefont {Leitner}\ and\ \citenamefont
  {Mosel}(2010{\natexlab{b}})}]{Leitner:2010jv}%
  \BibitemOpen
  \bibfield  {author} {\bibinfo {author} {\bibfnamefont {T.}~\bibnamefont
  {Leitner}}\ and\ \bibinfo {author} {\bibfnamefont {U.}~\bibnamefont
  {Mosel}},\ }\href {\doibase 10.1103/PhysRevC.82.035503} {\bibfield  {journal}
  {\bibinfo  {journal} {Phys.Rev.C}\ }\textbf {\bibinfo {volume} {82}},\
  \bibinfo {pages} {035503} (\bibinfo {year} {2010}{\natexlab{b}})},\ \Eprint
  {http://arxiv.org/abs/1006.2714} {arXiv:1006.2714 [nucl-th]} \BibitemShut
  {NoStop}%
\bibitem [{\citenamefont {Leitner}\ \emph {et~al.}(2010)\citenamefont
  {Leitner}, \citenamefont {Lalakulich}, \citenamefont {Buss}, \citenamefont
  {Mosel},\ and\ \citenamefont {Alvarez-Ruso}}]{Leitner:2009de}%
  \BibitemOpen
  \bibfield  {author} {\bibinfo {author} {\bibfnamefont {T.}~\bibnamefont
  {Leitner}}, \bibinfo {author} {\bibfnamefont {O.}~\bibnamefont {Lalakulich}},
  \bibinfo {author} {\bibfnamefont {O.}~\bibnamefont {Buss}}, \bibinfo {author}
  {\bibfnamefont {U.}~\bibnamefont {Mosel}}, \ and\ \bibinfo {author}
  {\bibfnamefont {L.}~\bibnamefont {Alvarez-Ruso}},\ }\href {\doibase
  10.1063/1.3399298} {\bibfield  {journal} {\bibinfo  {journal} {AIP
  Conf.Proc.}\ }\textbf {\bibinfo {volume} {1222}},\ \bibinfo {pages} {212}
  (\bibinfo {year} {2010})},\ \Eprint {http://arxiv.org/abs/0910.2835}
  {arXiv:0910.2835 [nucl-th]} \BibitemShut {NoStop}%
\bibitem [{\citenamefont {Lalakulich}\ \emph {et~al.}(2011)\citenamefont
  {Lalakulich}, \citenamefont {Gallmeister}, \citenamefont {Leitner},\ and\
  \citenamefont {Mosel}}]{Lalakulich:2011ne}%
  \BibitemOpen
  \bibfield  {author} {\bibinfo {author} {\bibfnamefont {O.}~\bibnamefont
  {Lalakulich}}, \bibinfo {author} {\bibfnamefont {K.}~\bibnamefont
  {Gallmeister}}, \bibinfo {author} {\bibfnamefont {T.}~\bibnamefont
  {Leitner}}, \ and\ \bibinfo {author} {\bibfnamefont {U.}~\bibnamefont
  {Mosel}},\ }\href {\doibase 10.1063/1.3661572} {\bibfield  {journal}
  {\bibinfo  {journal} {AIP Conf.Proc.}\ }\textbf {\bibinfo {volume} {1405}},\
  \bibinfo {pages} {127} (\bibinfo {year} {2011})},\ \Eprint
  {http://arxiv.org/abs/1107.5947} {arXiv:1107.5947 [nucl-th]} \BibitemShut
  {NoStop}%
\bibitem [{\citenamefont {Aguilar-Arevalo}\ \emph
  {et~al.}(2011{\natexlab{a}})\citenamefont {Aguilar-Arevalo} \emph
  {et~al.}}]{AguilarArevalo:2010xt}%
  \BibitemOpen
  \bibfield  {author} {\bibinfo {author} {\bibfnamefont {A.}~\bibnamefont
  {Aguilar-Arevalo}} \emph {et~al.} (\bibinfo {collaboration} {MiniBooNE
  Collaboration}),\ }\href {\doibase 10.1103/PhysRevD.83.052009} {\bibfield
  {journal} {\bibinfo  {journal} {Phys.Rev.D}\ }\textbf {\bibinfo {volume}
  {83}},\ \bibinfo {pages} {052009} (\bibinfo {year} {2011}{\natexlab{a}})},\
  \Eprint {http://arxiv.org/abs/1010.3264} {arXiv:1010.3264 [hep-ex]}
  \BibitemShut {NoStop}%
\bibitem [{\citenamefont {Aguilar-Arevalo}\ \emph
  {et~al.}(2011{\natexlab{b}})\citenamefont {Aguilar-Arevalo} \emph
  {et~al.}}]{AguilarArevalo:2010bm}%
  \BibitemOpen
  \bibfield  {author} {\bibinfo {author} {\bibfnamefont {A.}~\bibnamefont
  {Aguilar-Arevalo}} \emph {et~al.} (\bibinfo {collaboration} {MiniBooNE
  Collaboration}),\ }\href {\doibase 10.1103/PhysRevD.83.052007} {\bibfield
  {journal} {\bibinfo  {journal} {Phys.Rev.D}\ }\textbf {\bibinfo {volume}
  {83}},\ \bibinfo {pages} {052007} (\bibinfo {year} {2011}{\natexlab{b}})},\
  \Eprint {http://arxiv.org/abs/1011.3572} {arXiv:1011.3572 [hep-ex]}
  \BibitemShut {NoStop}%
\bibitem [{\citenamefont {Alvarez-Ruso}\ \emph {et~al.}(2009)\citenamefont
  {Alvarez-Ruso}, \citenamefont {Buss}, \citenamefont {Leitner},\ and\
  \citenamefont {Mosel}}]{AlvarezRuso:2009ad}%
  \BibitemOpen
  \bibfield  {author} {\bibinfo {author} {\bibfnamefont {L.}~\bibnamefont
  {Alvarez-Ruso}}, \bibinfo {author} {\bibfnamefont {O.}~\bibnamefont {Buss}},
  \bibinfo {author} {\bibfnamefont {T.}~\bibnamefont {Leitner}}, \ and\
  \bibinfo {author} {\bibfnamefont {U.}~\bibnamefont {Mosel}},\ }\href
  {\doibase 10.1063/1.3274146} {\bibfield  {journal} {\bibinfo  {journal} {AIP
  Conf. Proc.}\ }\textbf {\bibinfo {volume} {1189}},\ \bibinfo {pages} {151}
  (\bibinfo {year} {2009})},\ \Eprint {http://arxiv.org/abs/0909.5123}
  {arXiv:0909.5123 [nucl-th]} \BibitemShut {NoStop}%
\bibitem [{\citenamefont {Aguilar-Arevalo}\ \emph {et~al.}(2009)\citenamefont
  {Aguilar-Arevalo} \emph {et~al.}}]{AguilarArevalo:2008yp}%
  \BibitemOpen
  \bibfield  {author} {\bibinfo {author} {\bibfnamefont {A.}~\bibnamefont
  {Aguilar-Arevalo}} \emph {et~al.} (\bibinfo {collaboration} {MiniBooNE
  Collaboration}),\ }\href {\doibase 10.1103/PhysRevD.79.072002} {\bibfield
  {journal} {\bibinfo  {journal} {Phys.Rev.}\ }\textbf {\bibinfo {volume}
  {D79}},\ \bibinfo {pages} {072002} (\bibinfo {year} {2009})},\ \Eprint
  {http://arxiv.org/abs/0806.1449} {arXiv:0806.1449 [hep-ex]} \BibitemShut
  {NoStop}%
\bibitem [{\citenamefont {Bhattacharya}\ \emph {et~al.}(2011)\citenamefont
  {Bhattacharya}, \citenamefont {Hill},\ and\ \citenamefont
  {Paz}}]{Bhattacharya:2011ah}%
  \BibitemOpen
  \bibfield  {author} {\bibinfo {author} {\bibfnamefont {B.}~\bibnamefont
  {Bhattacharya}}, \bibinfo {author} {\bibfnamefont {R.~J.}\ \bibnamefont
  {Hill}}, \ and\ \bibinfo {author} {\bibfnamefont {G.}~\bibnamefont {Paz}},\
  }\href {\doibase 10.1103/PhysRevD.84.073006} {\bibfield  {journal} {\bibinfo
  {journal} {Phys.Rev.D}\ }\textbf {\bibinfo {volume} {84}},\ \bibinfo {pages}
  {073006} (\bibinfo {year} {2011})},\ \Eprint {http://arxiv.org/abs/1108.0423}
  {arXiv:1108.0423 [hep-ph]} \BibitemShut {NoStop}%
\bibitem [{\citenamefont {Bodek}\ \emph {et~al.}(2011)\citenamefont {Bodek},
  \citenamefont {Budd},\ and\ \citenamefont {Christie}}]{Bodek:2011ps}%
  \BibitemOpen
  \bibfield  {author} {\bibinfo {author} {\bibfnamefont {A.}~\bibnamefont
  {Bodek}}, \bibinfo {author} {\bibfnamefont {H.}~\bibnamefont {Budd}}, \ and\
  \bibinfo {author} {\bibfnamefont {M.}~\bibnamefont {Christie}},\ }\href
  {\doibase 10.1140/epjc/s10052-011-1726-y} {\bibfield  {journal} {\bibinfo
  {journal} {Eur.Phys.J.C}\ }\textbf {\bibinfo {volume} {71}},\ \bibinfo
  {pages} {1726} (\bibinfo {year} {2011})},\ \Eprint
  {http://arxiv.org/abs/1106.0340} {arXiv:1106.0340 [hep-ph]} \BibitemShut
  {NoStop}%
\bibitem [{\citenamefont {Sobczyk}(2012)}]{Sobczyk:2012ah}%
  \BibitemOpen
  \bibfield  {author} {\bibinfo {author} {\bibfnamefont {J.~T.}\ \bibnamefont
  {Sobczyk}},\ }\href {\doibase 10.1140/epjc/s10052-011-1850-8} {\bibfield
  {journal} {\bibinfo  {journal} {Eur.Phys.J.}\ }\textbf {\bibinfo {volume}
  {C72}},\ \bibinfo {pages} {1850} (\bibinfo {year} {2012})},\ \Eprint
  {http://arxiv.org/abs/1109.1081} {arXiv:1109.1081 [hep-ex]} \BibitemShut
  {NoStop}%
\bibitem [{\citenamefont {Meucci}\ \emph {et~al.}(2011)\citenamefont {Meucci},
  \citenamefont {Barbaro}, \citenamefont {Caballero}, \citenamefont {Giusti},\
  and\ \citenamefont {Udias}}]{Meucci:2011vd}%
  \BibitemOpen
  \bibfield  {author} {\bibinfo {author} {\bibfnamefont {A.}~\bibnamefont
  {Meucci}}, \bibinfo {author} {\bibfnamefont {M.}~\bibnamefont {Barbaro}},
  \bibinfo {author} {\bibfnamefont {J.}~\bibnamefont {Caballero}}, \bibinfo
  {author} {\bibfnamefont {C.}~\bibnamefont {Giusti}}, \ and\ \bibinfo {author}
  {\bibfnamefont {J.}~\bibnamefont {Udias}},\ }\href {\doibase
  10.1103/PhysRevLett.107.172501} {\bibfield  {journal} {\bibinfo  {journal}
  {Phys.Rev.Lett.}\ }\textbf {\bibinfo {volume} {107}},\ \bibinfo {pages}
  {172501} (\bibinfo {year} {2011})},\ \Eprint {http://arxiv.org/abs/1107.5145}
  {arXiv:1107.5145 [nucl-th]} \BibitemShut {NoStop}%
\bibitem [{\citenamefont {Bleve}\ \emph {et~al.}(2001)\citenamefont {Bleve},
  \citenamefont {Co}, \citenamefont {De~Mitri}, \citenamefont {Bernardini},
  \citenamefont {Mancarella} \emph {et~al.}}]{Bleve:2000hc}%
  \BibitemOpen
  \bibfield  {author} {\bibinfo {author} {\bibfnamefont {C.}~\bibnamefont
  {Bleve}}, \bibinfo {author} {\bibfnamefont {G.}~\bibnamefont {Co}}, \bibinfo
  {author} {\bibfnamefont {I.}~\bibnamefont {De~Mitri}}, \bibinfo {author}
  {\bibfnamefont {P.}~\bibnamefont {Bernardini}}, \bibinfo {author}
  {\bibfnamefont {G.}~\bibnamefont {Mancarella}},  \emph {et~al.},\ }\href
  {\doibase 10.1016/S0927-6505(01)00106-2} {\bibfield  {journal} {\bibinfo
  {journal} {Astropart.Phys.}\ }\textbf {\bibinfo {volume} {16}},\ \bibinfo
  {pages} {145} (\bibinfo {year} {2001})},\ \Eprint
  {http://arxiv.org/abs/nucl-th/0012015} {arXiv:nucl-th/0012015 [nucl-th]}
  \BibitemShut {NoStop}%
\bibitem [{\citenamefont {Alberico}\ \emph {et~al.}(1984)\citenamefont
  {Alberico}, \citenamefont {Ericson},\ and\ \citenamefont
  {Molinari}}]{Alberico:1983zg}%
  \BibitemOpen
  \bibfield  {author} {\bibinfo {author} {\bibfnamefont {W.}~\bibnamefont
  {Alberico}}, \bibinfo {author} {\bibfnamefont {M.}~\bibnamefont {Ericson}}, \
  and\ \bibinfo {author} {\bibfnamefont {A.}~\bibnamefont {Molinari}},\ }\href
  {\doibase 10.1016/0003-4916(84)90155-6} {\bibfield  {journal} {\bibinfo
  {journal} {Annals Phys.}\ }\textbf {\bibinfo {volume} {154}},\ \bibinfo
  {pages} {356} (\bibinfo {year} {1984})}\BibitemShut {NoStop}%
\bibitem [{\citenamefont {Kim}\ \emph {et~al.}(1995)\citenamefont {Kim},
  \citenamefont {Piekarewicz},\ and\ \citenamefont {Horowitz}}]{Kim:1994zea}%
  \BibitemOpen
  \bibfield  {author} {\bibinfo {author} {\bibfnamefont {H.-C.}\ \bibnamefont
  {Kim}}, \bibinfo {author} {\bibfnamefont {J.}~\bibnamefont {Piekarewicz}}, \
  and\ \bibinfo {author} {\bibfnamefont {C.}~\bibnamefont {Horowitz}},\ }\href
  {\doibase 10.1103/PhysRevC.51.2739} {\bibfield  {journal} {\bibinfo
  {journal} {Phys.Rev.C}\ }\textbf {\bibinfo {volume} {51}},\ \bibinfo {pages}
  {2739} (\bibinfo {year} {1995})},\ \Eprint
  {http://arxiv.org/abs/nucl-th/9412017} {arXiv:nucl-th/9412017 [nucl-th]}
  \BibitemShut {NoStop}%
\bibitem [{\citenamefont {Boffi}\ \emph {et~al.}(1996)\citenamefont {Boffi},
  \citenamefont {Giusti}, \citenamefont {Pacati},\ and\ \citenamefont
  {Radici}}]{Boffi:1994}%
  \BibitemOpen
  \bibfield  {author} {\bibinfo {author} {\bibfnamefont {S.}~\bibnamefont
  {Boffi}}, \bibinfo {author} {\bibfnamefont {C.}~\bibnamefont {Giusti}},
  \bibinfo {author} {\bibfnamefont {F.}~\bibnamefont {Pacati}}, \ and\ \bibinfo
  {author} {\bibfnamefont {M.}~\bibnamefont {Radici}},\ }\href@noop {} {\emph
  {\bibinfo {title} {Electromagnetic Response of Nuclei}}}\ (\bibinfo
  {publisher} {Clarendon Press},\ \bibinfo {address} {Oxford},\ \bibinfo {year}
  {1996})\BibitemShut {NoStop}%
\bibitem [{\citenamefont {Dekker}\ \emph {et~al.}(1994)\citenamefont {Dekker},
  \citenamefont {Brussaard},\ and\ \citenamefont {Tjon}}]{Dekker:1994yc}%
  \BibitemOpen
  \bibfield  {author} {\bibinfo {author} {\bibfnamefont {M.}~\bibnamefont
  {Dekker}}, \bibinfo {author} {\bibfnamefont {P.}~\bibnamefont {Brussaard}}, \
  and\ \bibinfo {author} {\bibfnamefont {J.}~\bibnamefont {Tjon}},\ }\href
  {\doibase 10.1103/PhysRevC.49.2650} {\bibfield  {journal} {\bibinfo
  {journal} {Phys.Rev.C}\ }\textbf {\bibinfo {volume} {49}},\ \bibinfo {pages}
  {2650} (\bibinfo {year} {1994})}\BibitemShut {NoStop}%
\bibitem [{\citenamefont {Delorme}\ and\ \citenamefont
  {Ericson}(1985)}]{Delorme:1985ps}%
  \BibitemOpen
  \bibfield  {author} {\bibinfo {author} {\bibfnamefont {J.}~\bibnamefont
  {Delorme}}\ and\ \bibinfo {author} {\bibfnamefont {M.}~\bibnamefont
  {Ericson}},\ }\href {\doibase 10.1016/0370-2693(85)91521-7} {\bibfield
  {journal} {\bibinfo  {journal} {Phys.Lett.}\ }\textbf {\bibinfo {volume}
  {B156}},\ \bibinfo {pages} {263} (\bibinfo {year} {1985})}\BibitemShut
  {NoStop}%
\bibitem [{\citenamefont {Marteau}(1999)}]{Marteau:1999kt}%
  \BibitemOpen
  \bibfield  {author} {\bibinfo {author} {\bibfnamefont {J.}~\bibnamefont
  {Marteau}},\ }\href {\doibase 10.1007/s100500050274} {\bibfield  {journal}
  {\bibinfo  {journal} {Eur.Phys.J.A}\ }\textbf {\bibinfo {volume} {5}},\
  \bibinfo {pages} {183} (\bibinfo {year} {1999})},\ \Eprint
  {http://arxiv.org/abs/hep-ph/9902210} {arXiv:hep-ph/9902210 [hep-ph]}
  \BibitemShut {NoStop}%
\bibitem [{\citenamefont {Martini}\ \emph {et~al.}(2009)\citenamefont
  {Martini}, \citenamefont {Ericson}, \citenamefont {Chanfray},\ and\
  \citenamefont {Marteau}}]{Martini:2009uj}%
  \BibitemOpen
  \bibfield  {author} {\bibinfo {author} {\bibfnamefont {M.}~\bibnamefont
  {Martini}}, \bibinfo {author} {\bibfnamefont {M.}~\bibnamefont {Ericson}},
  \bibinfo {author} {\bibfnamefont {G.}~\bibnamefont {Chanfray}}, \ and\
  \bibinfo {author} {\bibfnamefont {J.}~\bibnamefont {Marteau}},\ }\href
  {\doibase 10.1103/PhysRevC.80.065501} {\bibfield  {journal} {\bibinfo
  {journal} {Phys.Rev.C}\ }\textbf {\bibinfo {volume} {80}},\ \bibinfo {pages}
  {065501} (\bibinfo {year} {2009})},\ \Eprint {http://arxiv.org/abs/0910.2622}
  {arXiv:0910.2622 [nucl-th]} \BibitemShut {NoStop}%
\bibitem [{\citenamefont {Martini}\ \emph {et~al.}(2010)\citenamefont
  {Martini}, \citenamefont {Ericson}, \citenamefont {Chanfray},\ and\
  \citenamefont {Marteau}}]{Martini:2010ex}%
  \BibitemOpen
  \bibfield  {author} {\bibinfo {author} {\bibfnamefont {M.}~\bibnamefont
  {Martini}}, \bibinfo {author} {\bibfnamefont {M.}~\bibnamefont {Ericson}},
  \bibinfo {author} {\bibfnamefont {G.}~\bibnamefont {Chanfray}}, \ and\
  \bibinfo {author} {\bibfnamefont {J.}~\bibnamefont {Marteau}},\ }\href
  {\doibase 10.1103/PhysRevC.81.045502} {\bibfield  {journal} {\bibinfo
  {journal} {Phys.Rev.C}\ }\textbf {\bibinfo {volume} {81}},\ \bibinfo {pages}
  {045502} (\bibinfo {year} {2010})},\ \Eprint {http://arxiv.org/abs/1002.4538}
  {arXiv:1002.4538 [hep-ph]} \BibitemShut {NoStop}%
\bibitem [{\citenamefont {Martini}\ \emph {et~al.}(2011)\citenamefont
  {Martini}, \citenamefont {Ericson},\ and\ \citenamefont
  {Chanfray}}]{Martini:2011wp}%
  \BibitemOpen
  \bibfield  {author} {\bibinfo {author} {\bibfnamefont {M.}~\bibnamefont
  {Martini}}, \bibinfo {author} {\bibfnamefont {M.}~\bibnamefont {Ericson}}, \
  and\ \bibinfo {author} {\bibfnamefont {G.}~\bibnamefont {Chanfray}},\ }\href
  {\doibase 10.1103/PhysRevC.84.055502} {\bibfield  {journal} {\bibinfo
  {journal} {Phys.Rev.}\ }\textbf {\bibinfo {volume} {C84}},\ \bibinfo {pages}
  {055502} (\bibinfo {year} {2011})},\ \Eprint {http://arxiv.org/abs/1110.0221}
  {arXiv:1110.0221 [nucl-th]} \BibitemShut {NoStop}%
\bibitem [{\citenamefont {Nieves}\ \emph {et~al.}(2011)\citenamefont {Nieves},
  \citenamefont {Ruiz~Simo},\ and\ \citenamefont
  {Vicente~Vacas}}]{Nieves:2011pp}%
  \BibitemOpen
  \bibfield  {author} {\bibinfo {author} {\bibfnamefont {J.}~\bibnamefont
  {Nieves}}, \bibinfo {author} {\bibfnamefont {I.}~\bibnamefont {Ruiz~Simo}}, \
  and\ \bibinfo {author} {\bibfnamefont {M.}~\bibnamefont {Vicente~Vacas}},\
  }\href {\doibase 10.1103/PhysRevC.83.045501} {\bibfield  {journal} {\bibinfo
  {journal} {Phys.Rev.C}\ }\textbf {\bibinfo {volume} {83}},\ \bibinfo {pages}
  {045501} (\bibinfo {year} {2011})},\ \Eprint {http://arxiv.org/abs/1102.2777}
  {arXiv:1102.2777 [hep-ph]} \BibitemShut {NoStop}%
\bibitem [{\citenamefont {Boyd}\ \emph {et~al.}(2009)\citenamefont {Boyd},
  \citenamefont {Dytman}, \citenamefont {Hernandez}, \citenamefont {Sobczyk},\
  and\ \citenamefont {Tacik}}]{Boyd:2009zz}%
  \BibitemOpen
  \bibfield  {author} {\bibinfo {author} {\bibfnamefont {S.}~\bibnamefont
  {Boyd}}, \bibinfo {author} {\bibfnamefont {S.}~\bibnamefont {Dytman}},
  \bibinfo {author} {\bibfnamefont {E.}~\bibnamefont {Hernandez}}, \bibinfo
  {author} {\bibfnamefont {J.}~\bibnamefont {Sobczyk}}, \ and\ \bibinfo
  {author} {\bibfnamefont {R.}~\bibnamefont {Tacik}},\ }\href {\doibase
  10.1063/1.3274191} {\bibfield  {journal} {\bibinfo  {journal} {AIP
  Conf.Proc.}\ }\textbf {\bibinfo {volume} {1189}},\ \bibinfo {pages} {60}
  (\bibinfo {year} {2009})}\BibitemShut {NoStop}%
\bibitem [{\citenamefont {Nieves}\ and\ \citenamefont
  {Vicente~Vacas}(2011)}]{Nieves-Vicente:2011}%
  \BibitemOpen
  \bibfield  {author} {\bibinfo {author} {\bibfnamefont {J.}~\bibnamefont
  {Nieves}}\ and\ \bibinfo {author} {\bibfnamefont {M.}~\bibnamefont
  {Vicente~Vacas}},\ }\href@noop {} {}\bibinfo {howpublished} {private
  communication} (\bibinfo {year} {2011})\BibitemShut {NoStop}%
\bibitem [{\citenamefont {Nieves}\ \emph
  {et~al.}(2012{\natexlab{a}})\citenamefont {Nieves}, \citenamefont {Simo},\
  and\ \citenamefont {Vacas}}]{Nieves:2011yp}%
  \BibitemOpen
  \bibfield  {author} {\bibinfo {author} {\bibfnamefont {J.}~\bibnamefont
  {Nieves}}, \bibinfo {author} {\bibfnamefont {I.}~\bibnamefont {Simo}}, \ and\
  \bibinfo {author} {\bibfnamefont {M.}~\bibnamefont {Vacas}},\ }\href
  {\doibase 10.1016/j.physletb.2011.11.061} {\bibfield  {journal} {\bibinfo
  {journal} {Phys.Lett.}\ }\textbf {\bibinfo {volume} {B707}},\ \bibinfo
  {pages} {72} (\bibinfo {year} {2012}{\natexlab{a}})},\ \Eprint
  {http://arxiv.org/abs/1106.5374} {arXiv:1106.5374 [hep-ph]} \BibitemShut
  {NoStop}%
\bibitem [{\citenamefont {Gil}\ \emph {et~al.}(1997{\natexlab{a}})\citenamefont
  {Gil}, \citenamefont {Nieves},\ and\ \citenamefont {Oset}}]{Gil:1997bm}%
  \BibitemOpen
  \bibfield  {author} {\bibinfo {author} {\bibfnamefont {A.}~\bibnamefont
  {Gil}}, \bibinfo {author} {\bibfnamefont {J.}~\bibnamefont {Nieves}}, \ and\
  \bibinfo {author} {\bibfnamefont {E.}~\bibnamefont {Oset}},\ }\href {\doibase
  10.1016/S0375-9474(97)00513-7} {\bibfield  {journal} {\bibinfo  {journal}
  {Nucl.Phys.A}\ }\textbf {\bibinfo {volume} {627}},\ \bibinfo {pages} {543}
  (\bibinfo {year} {1997}{\natexlab{a}})},\ \Eprint
  {http://arxiv.org/abs/nucl-th/9711009} {arXiv:nucl-th/9711009 [nucl-th]}
  \BibitemShut {NoStop}%
\bibitem [{\citenamefont {Gil}\ \emph {et~al.}(1997{\natexlab{b}})\citenamefont
  {Gil}, \citenamefont {Nieves},\ and\ \citenamefont {Oset}}]{Gil:1997jg}%
  \BibitemOpen
  \bibfield  {author} {\bibinfo {author} {\bibfnamefont {A.}~\bibnamefont
  {Gil}}, \bibinfo {author} {\bibfnamefont {J.}~\bibnamefont {Nieves}}, \ and\
  \bibinfo {author} {\bibfnamefont {E.}~\bibnamefont {Oset}},\ }\href {\doibase
  10.1016/S0375-9474(97)00515-0} {\bibfield  {journal} {\bibinfo  {journal}
  {Nucl.Phys.A}\ }\textbf {\bibinfo {volume} {627}},\ \bibinfo {pages} {599}
  (\bibinfo {year} {1997}{\natexlab{b}})},\ \Eprint
  {http://arxiv.org/abs/nucl-th/9710070} {arXiv:nucl-th/9710070 [nucl-th]}
  \BibitemShut {NoStop}%
\bibitem [{\citenamefont {Amaro}\ \emph
  {et~al.}(2011{\natexlab{a}})\citenamefont {Amaro}, \citenamefont {Barbaro},
  \citenamefont {Caballero}, \citenamefont {Donnelly},\ and\ \citenamefont
  {Williamson}}]{Amaro:2010sd}%
  \BibitemOpen
  \bibfield  {author} {\bibinfo {author} {\bibfnamefont {J.}~\bibnamefont
  {Amaro}}, \bibinfo {author} {\bibfnamefont {M.}~\bibnamefont {Barbaro}},
  \bibinfo {author} {\bibfnamefont {J.}~\bibnamefont {Caballero}}, \bibinfo
  {author} {\bibfnamefont {T.}~\bibnamefont {Donnelly}}, \ and\ \bibinfo
  {author} {\bibfnamefont {C.}~\bibnamefont {Williamson}},\ }\href {\doibase
  10.1016/j.physletb.2010.12.007} {\bibfield  {journal} {\bibinfo  {journal}
  {Phys.Lett.B}\ }\textbf {\bibinfo {volume} {696}},\ \bibinfo {pages} {151}
  (\bibinfo {year} {2011}{\natexlab{a}})},\ \Eprint
  {http://arxiv.org/abs/1010.1708} {arXiv:1010.1708 [nucl-th]} \BibitemShut
  {NoStop}%
\bibitem [{\citenamefont {Amaro}\ \emph
  {et~al.}(2011{\natexlab{b}})\citenamefont {Amaro}, \citenamefont {Barbaro},
  \citenamefont {Caballero}, \citenamefont {Donnelly},\ and\ \citenamefont
  {Udias}}]{Amaro:2011qb}%
  \BibitemOpen
  \bibfield  {author} {\bibinfo {author} {\bibfnamefont {J.}~\bibnamefont
  {Amaro}}, \bibinfo {author} {\bibfnamefont {M.}~\bibnamefont {Barbaro}},
  \bibinfo {author} {\bibfnamefont {J.}~\bibnamefont {Caballero}}, \bibinfo
  {author} {\bibfnamefont {T.}~\bibnamefont {Donnelly}}, \ and\ \bibinfo
  {author} {\bibfnamefont {J.}~\bibnamefont {Udias}},\ }\href {\doibase
  10.1103/PhysRevD.84.033004} {\bibfield  {journal} {\bibinfo  {journal}
  {Phys.Rev.D}\ }\textbf {\bibinfo {volume} {84}},\ \bibinfo {pages} {033004}
  (\bibinfo {year} {2011}{\natexlab{b}})},\ \Eprint
  {http://arxiv.org/abs/1104.5446} {arXiv:1104.5446 [nucl-th]} \BibitemShut
  {NoStop}%
\bibitem [{\citenamefont {Amaro}\ \emph
  {et~al.}(2011{\natexlab{c}})\citenamefont {Amaro}, \citenamefont {Barbaro},
  \citenamefont {Caballero},\ and\ \citenamefont {Donnelly}}]{Amaro:2011aa}%
  \BibitemOpen
  \bibfield  {author} {\bibinfo {author} {\bibfnamefont {J.}~\bibnamefont
  {Amaro}}, \bibinfo {author} {\bibfnamefont {M.}~\bibnamefont {Barbaro}},
  \bibinfo {author} {\bibfnamefont {J.}~\bibnamefont {Caballero}}, \ and\
  \bibinfo {author} {\bibfnamefont {T.}~\bibnamefont {Donnelly}},\ }\href@noop
  {} {\  (\bibinfo {year} {2011}{\natexlab{c}})},\ \Eprint
  {http://arxiv.org/abs/1112.2123} {arXiv:1112.2123 [nucl-th]} \BibitemShut
  {NoStop}%
\bibitem [{\citenamefont {Martini}(2011)}]{Martini:2011ui}%
  \BibitemOpen
  \bibfield  {author} {\bibinfo {author} {\bibfnamefont {M.}~\bibnamefont
  {Martini}},\ }\href@noop {} {\  (\bibinfo {year} {2011})},\ \Eprint
  {http://arxiv.org/abs/1110.5895} {arXiv:1110.5895 [hep-ph]} \BibitemShut
  {NoStop}%
\bibitem [{\citenamefont {Alvarez-Ruso}(2011)}]{Alvarez-Ruso:2011}%
  \BibitemOpen
  \bibfield  {author} {\bibinfo {author} {\bibfnamefont {L.}~\bibnamefont
  {Alvarez-Ruso}},\ }\href {\doibase 10.1063/1.3661561} {\bibfield  {journal}
  {\bibinfo  {journal} {AIP Conf.Proc.}\ }\textbf {\bibinfo {volume} {1405}},\
  \bibinfo {pages} {71} (\bibinfo {year} {2011})}\BibitemShut {NoStop}%
\bibitem [{\citenamefont {Ankowski}\ \emph {et~al.}(2010)\citenamefont
  {Ankowski}, \citenamefont {Benhar},\ and\ \citenamefont
  {Farina}}]{Ankowski:2010yh}%
  \BibitemOpen
  \bibfield  {author} {\bibinfo {author} {\bibfnamefont {A.~M.}\ \bibnamefont
  {Ankowski}}, \bibinfo {author} {\bibfnamefont {O.}~\bibnamefont {Benhar}}, \
  and\ \bibinfo {author} {\bibfnamefont {N.}~\bibnamefont {Farina}},\ }\href
  {\doibase 10.1103/PhysRevD.82.013002} {\bibfield  {journal} {\bibinfo
  {journal} {Phys.Rev.D}\ }\textbf {\bibinfo {volume} {82}},\ \bibinfo {pages}
  {013002} (\bibinfo {year} {2010})},\ \Eprint {http://arxiv.org/abs/1001.0481}
  {arXiv:1001.0481 [nucl-th]} \BibitemShut {NoStop}%
\bibitem [{\citenamefont {Buss}\ \emph {et~al.}(2012)\citenamefont {Buss},
  \citenamefont {Gaitanos}, \citenamefont {Gallmeister}, \citenamefont {van
  Hees}, \citenamefont {Kaskulov} \emph {et~al.}}]{Buss:2011mx}%
  \BibitemOpen
  \bibfield  {author} {\bibinfo {author} {\bibfnamefont {O.}~\bibnamefont
  {Buss}}, \bibinfo {author} {\bibfnamefont {T.}~\bibnamefont {Gaitanos}},
  \bibinfo {author} {\bibfnamefont {K.}~\bibnamefont {Gallmeister}}, \bibinfo
  {author} {\bibfnamefont {H.}~\bibnamefont {van Hees}}, \bibinfo {author}
  {\bibfnamefont {M.}~\bibnamefont {Kaskulov}},  \emph {et~al.},\ }\href
  {\doibase 10.1016/j.physrep.2011.12.001} {\bibfield  {journal} {\bibinfo
  {journal} {Phys.Rept.}\ }\textbf {\bibinfo {volume} {512}},\ \bibinfo {pages}
  {1} (\bibinfo {year} {2012})},\ \Eprint {http://arxiv.org/abs/1106.1344}
  {arXiv:1106.1344 [hep-ph]} \BibitemShut {NoStop}%
\bibitem [{\citenamefont {Kadanoff}\ and\ \citenamefont
  {Baym}(1962)}]{Kad-Baym:1962}%
  \BibitemOpen
  \bibfield  {author} {\bibinfo {author} {\bibfnamefont {L.}~\bibnamefont
  {Kadanoff}}\ and\ \bibinfo {author} {\bibfnamefont {G.}~\bibnamefont
  {Baym}},\ }\href@noop {} {\emph {\bibinfo {title} {Quantum statistical
  mechanics}}}\ (\bibinfo  {publisher} {Benjamin},\ \bibinfo {address} {New
  York},\ \bibinfo {year} {1962})\BibitemShut {NoStop}%
\bibitem [{\citenamefont {Botermans}\ and\ \citenamefont
  {Malfliet}(1990)}]{Botermans:1990qi}%
  \BibitemOpen
  \bibfield  {author} {\bibinfo {author} {\bibfnamefont {W.}~\bibnamefont
  {Botermans}}\ and\ \bibinfo {author} {\bibfnamefont {R.}~\bibnamefont
  {Malfliet}},\ }\href {\doibase 10.1016/0370-1573(90)90174-Z} {\bibfield
  {journal} {\bibinfo  {journal} {Phys.Rept.}\ }\textbf {\bibinfo {volume}
  {198}},\ \bibinfo {pages} {115} (\bibinfo {year} {1990})}\BibitemShut
  {NoStop}%
\bibitem [{\citenamefont {Leitner}\ \emph {et~al.}(2009)\citenamefont
  {Leitner}, \citenamefont {Buss},\ and\ \citenamefont
  {Mosel}}]{Leitner:2009ke}%
  \BibitemOpen
  \bibfield  {author} {\bibinfo {author} {\bibfnamefont {T.}~\bibnamefont
  {Leitner}}, \bibinfo {author} {\bibfnamefont {O.}~\bibnamefont {Buss}}, \
  and\ \bibinfo {author} {\bibfnamefont {U.}~\bibnamefont {Mosel}},\
  }\href@noop {} {\bibfield  {journal} {\bibinfo  {journal} {Acta
  Phys.Polon.B}\ }\textbf {\bibinfo {volume} {40}},\ \bibinfo {pages} {2585}
  (\bibinfo {year} {2009})},\ \Eprint {http://arxiv.org/abs/0905.1644}
  {arXiv:0905.1644 [nucl-th]} \BibitemShut {NoStop}%
\bibitem [{\citenamefont {Leitner}\ \emph
  {et~al.}(2006{\natexlab{a}})\citenamefont {Leitner}, \citenamefont
  {Alvarez-Ruso},\ and\ \citenamefont {Mosel}}]{Leitner:2006ww}%
  \BibitemOpen
  \bibfield  {author} {\bibinfo {author} {\bibfnamefont {T.}~\bibnamefont
  {Leitner}}, \bibinfo {author} {\bibfnamefont {L.}~\bibnamefont
  {Alvarez-Ruso}}, \ and\ \bibinfo {author} {\bibfnamefont {U.}~\bibnamefont
  {Mosel}},\ }\href {\doibase 10.1103/PhysRevC.73.065502} {\bibfield  {journal}
  {\bibinfo  {journal} {Phys. Rev.}\ }\textbf {\bibinfo {volume} {C73}},\
  \bibinfo {pages} {065502} (\bibinfo {year} {2006}{\natexlab{a}})},\ \Eprint
  {http://arxiv.org/abs/nucl-th/0601103} {arXiv:nucl-th/0601103} \BibitemShut
  {NoStop}%
\bibitem [{\citenamefont {Leitner}\ \emph
  {et~al.}(2006{\natexlab{b}})\citenamefont {Leitner}, \citenamefont
  {Alvarez-Ruso},\ and\ \citenamefont {Mosel}}]{Leitner:2006sp}%
  \BibitemOpen
  \bibfield  {author} {\bibinfo {author} {\bibfnamefont {T.}~\bibnamefont
  {Leitner}}, \bibinfo {author} {\bibfnamefont {L.}~\bibnamefont
  {Alvarez-Ruso}}, \ and\ \bibinfo {author} {\bibfnamefont {U.}~\bibnamefont
  {Mosel}},\ }\href {\doibase 10.1103/PhysRevC.74.065502} {\bibfield  {journal}
  {\bibinfo  {journal} {Phys. Rev.}\ }\textbf {\bibinfo {volume} {C74}},\
  \bibinfo {pages} {065502} (\bibinfo {year} {2006}{\natexlab{b}})},\ \Eprint
  {http://arxiv.org/abs/nucl-th/0606058} {arXiv:nucl-th/0606058} \BibitemShut
  {NoStop}%
\bibitem [{\citenamefont {Nieves}\ \emph {et~al.}(2004)\citenamefont {Nieves},
  \citenamefont {Amaro},\ and\ \citenamefont {Valverde}}]{Nieves:2004wx}%
  \BibitemOpen
  \bibfield  {author} {\bibinfo {author} {\bibfnamefont {J.}~\bibnamefont
  {Nieves}}, \bibinfo {author} {\bibfnamefont {J.~E.}\ \bibnamefont {Amaro}}, \
  and\ \bibinfo {author} {\bibfnamefont {M.}~\bibnamefont {Valverde}},\ }\href
  {\doibase 10.1103/PhysRevC.70.055503, 10.1103/PhysRevC.72.019902} {\bibfield
  {journal} {\bibinfo  {journal} {Phys.Rev.C}\ }\textbf {\bibinfo {volume}
  {70}},\ \bibinfo {pages} {055503} (\bibinfo {year} {2004})},\ \Eprint
  {http://arxiv.org/abs/nucl-th/0408005} {arXiv:nucl-th/0408005 [nucl-th]}
  \BibitemShut {NoStop}%
\bibitem [{\citenamefont {Gottfried}(1957)}]{Gottfried:1975}%
  \BibitemOpen
  \bibfield  {author} {\bibinfo {author} {\bibfnamefont {K.}~\bibnamefont
  {Gottfried}},\ }\href {\doibase 10.1016/0029-5582(58)90056-7} {\bibfield
  {journal} {\bibinfo  {journal} {Nucl.Phys.}\ }\textbf {\bibinfo {volume}
  {5}},\ \bibinfo {pages} {557} (\bibinfo {year} {1957})}\BibitemShut {NoStop}%
\bibitem [{\citenamefont {Giusti}\ and\ \citenamefont
  {Pacati}(1991)}]{Giusti:1991bs}%
  \BibitemOpen
  \bibfield  {author} {\bibinfo {author} {\bibfnamefont {C.}~\bibnamefont
  {Giusti}}\ and\ \bibinfo {author} {\bibfnamefont {F.}~\bibnamefont
  {Pacati}},\ }\href {\doibase 10.1016/0375-9474(91)90476-M} {\bibfield
  {journal} {\bibinfo  {journal} {Nucl.Phys.A}\ }\textbf {\bibinfo {volume}
  {535}},\ \bibinfo {pages} {573} (\bibinfo {year} {1991})}\BibitemShut
  {NoStop}%
\bibitem [{\citenamefont {Giusti}\ and\ \citenamefont
  {Pacati}(1994)}]{Giusti:1993}%
  \BibitemOpen
  \bibfield  {author} {\bibinfo {author} {\bibfnamefont {C.}~\bibnamefont
  {Giusti}}\ and\ \bibinfo {author} {\bibfnamefont {F.}~\bibnamefont
  {Pacati}},\ }\href@noop {} {\bibfield  {journal} {\bibinfo  {journal}
  {Nucl.Phys.A}\ }\textbf {\bibinfo {volume} {571}},\ \bibinfo {pages} {694}
  (\bibinfo {year} {1994})}\BibitemShut {NoStop}%
\bibitem [{\citenamefont {Van~der Sluys}\ \emph {et~al.}(1995)\citenamefont
  {Van~der Sluys}, \citenamefont {Ryckebusch},\ and\ \citenamefont
  {Waroquier}}]{VanderSluys:1995rp}%
  \BibitemOpen
  \bibfield  {author} {\bibinfo {author} {\bibfnamefont {V.}~\bibnamefont
  {Van~der Sluys}}, \bibinfo {author} {\bibfnamefont {J.}~\bibnamefont
  {Ryckebusch}}, \ and\ \bibinfo {author} {\bibfnamefont {M.}~\bibnamefont
  {Waroquier}},\ }\href {\doibase 10.1103/PhysRevC.51.2664} {\bibfield
  {journal} {\bibinfo  {journal} {Phys.Rev.C}\ }\textbf {\bibinfo {volume}
  {51}},\ \bibinfo {pages} {2664} (\bibinfo {year} {1995})},\ \Eprint
  {http://arxiv.org/abs/nucl-th/9503008} {arXiv:nucl-th/9503008 [nucl-th]}
  \BibitemShut {NoStop}%
\bibitem [{\citenamefont {Ryckebusch}(1996)}]{Ryckebusch:1996wc}%
  \BibitemOpen
  \bibfield  {author} {\bibinfo {author} {\bibfnamefont {J.}~\bibnamefont
  {Ryckebusch}},\ }\href {\doibase 10.1016/0370-2693(96)00725-3} {\bibfield
  {journal} {\bibinfo  {journal} {Phys.Lett.B}\ }\textbf {\bibinfo {volume}
  {383}},\ \bibinfo {pages} {1} (\bibinfo {year} {1996})},\ \Eprint
  {http://arxiv.org/abs/nucl-th/9605043} {arXiv:nucl-th/9605043 [nucl-th]}
  \BibitemShut {NoStop}%
\bibitem [{\citenamefont {Ruiz~Simo}\ \emph {et~al.}(2014)\citenamefont
  {Ruiz~Simo}, \citenamefont {Albertus}, \citenamefont {Amaro}, \citenamefont
  {Barbaro}, \citenamefont {Caballero} \emph {et~al.}}]{Simo:2014wka}%
  \BibitemOpen
  \bibfield  {author} {\bibinfo {author} {\bibfnamefont {I.}~\bibnamefont
  {Ruiz~Simo}}, \bibinfo {author} {\bibfnamefont {C.}~\bibnamefont {Albertus}},
  \bibinfo {author} {\bibfnamefont {J.}~\bibnamefont {Amaro}}, \bibinfo
  {author} {\bibfnamefont {M.}~\bibnamefont {Barbaro}}, \bibinfo {author}
  {\bibfnamefont {J.}~\bibnamefont {Caballero}},  \emph {et~al.},\ }\href@noop
  {} {\  (\bibinfo {year} {2014})},\ \Eprint {http://arxiv.org/abs/1405.4280}
  {arXiv:1405.4280 [nucl-th]} \BibitemShut {NoStop}%
\bibitem [{\citenamefont {Boffi}\ \emph {et~al.}(1993)\citenamefont {Boffi},
  \citenamefont {Giusti},\ and\ \citenamefont {Pacati}}]{Boffi:1993gs}%
  \BibitemOpen
  \bibfield  {author} {\bibinfo {author} {\bibfnamefont {S.}~\bibnamefont
  {Boffi}}, \bibinfo {author} {\bibfnamefont {C.}~\bibnamefont {Giusti}}, \
  and\ \bibinfo {author} {\bibfnamefont {F.}~\bibnamefont {Pacati}},\ }\href
  {\doibase 10.1016/0370-1573(93)90132-W} {\bibfield  {journal} {\bibinfo
  {journal} {Phys.Rept.}\ }\textbf {\bibinfo {volume} {226}},\ \bibinfo {pages}
  {1} (\bibinfo {year} {1993})}\BibitemShut {NoStop}%
\bibitem [{\citenamefont {Nieves}\ \emph
  {et~al.}(2012{\natexlab{b}})\citenamefont {Nieves}, \citenamefont {Sanchez},
  \citenamefont {Ruiz~Simo},\ and\ \citenamefont {Vacas}}]{Nieves:2012yz}%
  \BibitemOpen
  \bibfield  {author} {\bibinfo {author} {\bibfnamefont {J.}~\bibnamefont
  {Nieves}}, \bibinfo {author} {\bibfnamefont {F.}~\bibnamefont {Sanchez}},
  \bibinfo {author} {\bibfnamefont {I.}~\bibnamefont {Ruiz~Simo}}, \ and\
  \bibinfo {author} {\bibfnamefont {M.}~\bibnamefont {Vacas}},\ }\href
  {\doibase 10.1103/PhysRevD.85.113008} {\bibfield  {journal} {\bibinfo
  {journal} {Phys.Rev.}\ }\textbf {\bibinfo {volume} {D85}},\ \bibinfo {pages}
  {113008} (\bibinfo {year} {2012}{\natexlab{b}})},\ \Eprint
  {http://arxiv.org/abs/1204.5404} {arXiv:1204.5404 [hep-ph]} \BibitemShut
  {NoStop}%
\bibitem [{\citenamefont {Martini}\ \emph {et~al.}(2012)\citenamefont
  {Martini}, \citenamefont {Ericson},\ and\ \citenamefont
  {Chanfray}}]{Martini:2012fa}%
  \BibitemOpen
  \bibfield  {author} {\bibinfo {author} {\bibfnamefont {M.}~\bibnamefont
  {Martini}}, \bibinfo {author} {\bibfnamefont {M.}~\bibnamefont {Ericson}}, \
  and\ \bibinfo {author} {\bibfnamefont {G.}~\bibnamefont {Chanfray}},\ }\href
  {\doibase 10.1103/PhysRevD.85.093012} {\bibfield  {journal} {\bibinfo
  {journal} {Phys.Rev.}\ }\textbf {\bibinfo {volume} {D85}},\ \bibinfo {pages}
  {093012} (\bibinfo {year} {2012})},\ \Eprint {http://arxiv.org/abs/1202.4745}
  {arXiv:1202.4745 [hep-ph]} \BibitemShut {NoStop}%
\end{thebibliography}%

\end{document}